\documentclass[twocolumn,showpacs,preprintnumbers,amsmath,amssymb,prl]{revtex4}

\usepackage{graphicx}
\usepackage{bm}

\begin{document}

\title{Breakdown of The Excess Entropy Scaling for the Systems with Thermodynamic Anomalies}

\author{Yu. D. Fomin}
\affiliation{Institute for High Pressure Physics, Russian Academy
of Sciences, Troitsk 142190, Moscow Region, Russia}

\author{N. V. Gribova}
\affiliation{Institute for High Pressure Physics, Russian Academy
of Sciences, Troitsk 142190, Moscow Region, Russia}

\author{V. N. Ryzhov}
\affiliation{Institute for High Pressure Physics, Russian Academy
of Sciences, Troitsk 142190, Moscow Region, Russia}

\date{\today}

\begin{abstract}
This articles presents a simulation study of the applicability of
the Rosenfeld entropy scaling to the systems which can not be
approximated by effective hard spheres. Three systems are studied:
Herzian spheres, Gauss Core Model and soft repulsive shoulder
potential. These systems demonstrate the diffusion anomalies at
low temperatures: the diffusion increases with increasing density
or pressure. It is shown that for the first two systems which
belong to the class of bounded potentials the Rosenfeld scaling
formula is valid only in the infinite temperature limit where
there are no anomalies. For the soft repulsive shoulder the
scaling formula is valid already at sufficiently low temperatures,
however, out of the anomaly range.
\end{abstract}

 \maketitle

\section{I. Introduction}

It is well known that some liquids (for example, water, silica,
silicon, carbon, and phosphorus) show anomalous behavior in the
vicinity of their freezing lines \cite{deben2003,bul2002,
angel2004,book,book1,deben2001,netz,stanley1,ad1,ad2,ad3,ad4,ad5,
ad6,ad7,ad8,errington,mittal}.
The water phase diagrams have regions where a thermal expansion
coefficient is negative (density anomaly), a self-diffusivity
increases upon pressuring (diffusion anomaly), and the structural
order of the system decreases upon compression (structural
anomaly) \cite{deben2001,netz}. The regions where these anomalies
take place form nested domains in the density-temperature
\cite{deben2001} (or pressure-temperature \cite{netz}) planes: the
density anomaly region is inside the diffusion anomaly domain, and
both of these anomalous regions are inside the broader
structurally anomalous region. It is natural to relate this kind
of behavior with the orientational anisotropy of the potentials,
however, there are a number of studies which demonstrate the
waterlike anomalies in fluids that interact through spherically
symmetric potentials
\cite{8,9,10,11,12,13,14,15,16,17,18,19,20,21,22,23,24,25,26,
barb2008-1,barb2008,FFGRS2008,RS2002,RS2003,FRT2006,GFFR2009}.

It was shown \cite{errington,mittal} that the thermodynamic and
kinetic anomalies may be linked through excess entropy. In
particular, in Refs. \onlinecite{errington,mittal} the authors
propose that entropy scaling relations developed by Rosenfeld
\cite{ros1,ros2} can be used to describe the the regions of
diffusivity anomaly.

Rosenfeld based his arguments on the approximations of liquid by
an effective hard spheres system. In this approach the kinetic
coefficients are expressed in reduced units based on the mean
length related to density of the system $d=\rho^{-1/3}$ and
thermal velocity $v_{th}=(k_BT/m)^{1/2}$. The reduced diffusion
coefficient $D^*$, viscosity $\eta^*$ and thermal conductivity
$\kappa^*$ are written in the form

\begin{equation}
 D^*=D \frac{\rho ^{1/3}}{(k_BT/m)^{1/3}}
\end{equation}

\begin{equation}
 \eta^*=\eta \frac{\rho^{-2/3}}{(mk_BT)^{1/2}}
\end{equation}

\begin{equation}
 \kappa^*=\kappa \frac{\rho^{-2/3}}{k_B(k_BT/m)^{1/2}}
\end{equation}

Rosenfeld suggested that the reduced transport coefficients can be
connected to the excess entropy of the system
$S_{ex}=(S-S_{id})/{(Nk_B)}$ through the formula

\begin{equation}
  X=a_X \cdot e^{b_X S_{ex}},
\end{equation}

where $X$ is the transport coefficient, and $a_X$ and $b_X$ are
constants which depend on the studying property \cite{ros2}.
Interestingly the coefficients $a$ and $b$ show extremely weak
dependence on the material and can be considered as universal.

Another expression for relating diffusion coefficient to the
excess entropy was suggested by Dzugutov \cite{dzugmat}. In this
approach the natural parameters of the system were chosen to be
the particle diameter $\sigma$ and the Enskog collision frequency
$\Gamma_E=4 \sigma ^2 g(\sigma) \rho \surd {\frac{\pi k_BT}{m}}$,
where $g(\sigma)$ is the value of radial distribution function at
contact. In case of continuous potentials the value of $\sigma$
corresponds to the distance of the first maximum of radial
distribution function. Defining the reduced diffusion coefficient
as $D_D^*=\frac{D}{\Gamma_E \sigma ^2}$. Dzugutov suggested the
following formula for it

\begin{equation}
  D_D^*=0.049 e^{s_2},
\end{equation}

where $s_2$ is a pair contribution to the excess entropy
\cite{dzugmat}. It was shown that this relation holds for many
simple liquids. At the same time this equation is not strictly
valid for liquid metals. In the work \cite{dzugbreak} it was shown
that in this case it is necessary to replace the pair entropy
$s_2$ by the full excess entropy $S_{ex}$. It was also shown that
Dzugutov formula does not work for silica modelled with an angular
dependent potential \cite{dzugbreak}. It allows to say that
Rosenfeld relation is more general.

Remember that original idea underlying the Rosenfeld relation is
to refer the system under investigation to the hard spheres
system. In this respect it is interesting whether the Rosenfeld
scaling relation is valid for the systems essentially different
from the hard spheres. One of the examples of such systems is the
system with potentials with negative curvature \cite{8,9}. It was
shown in many publications that the behavior of such systems is
very complex
\cite{camp,malpel,norkaw,FRT2006,FFGRS2008,cst01,GFFR2009,fominpot,
malnegcurv,buld2009}. In particular such systems can form
complicated structures, like cluster liquids or different crystal
phases. They can demonstrate maximum on the melting line and
reentrant melting and many other unusual properties. In particular
systems with negative curvature can demonstrate anomalous behavior
\cite{GFFR2009,15,errington,20,bar2,19,india,barb2008-1}.

It was suggested that the Rosenfeld relations can hold even in the
case of anomalous diffusion \cite{errington,mittal}. For example,
in the paper \cite{mittal} the dependence of both excess entropy
and diffusion coefficient on density are reported for the
core-softening potential  that consists of a combination of a
Lennard-Jones potential  plus a Gaussian well. This potential can
represent a whole family of two length scales intermolecular
interactions, from a deep double-well potential to a repulsive
shoulder. Accordingly to these dependencies both excess entropy
and diffusion have non monotonic behavior which allows to preserve
exponential dependence of the diffusion coefficient on the excess
entropy. It means that the thermodynamically anomalous regions are
characterized by anomalous behavior of the excess entropy which
induces anomalous diffusion as well.

Another example of systems which can not be approximated by a hard
sphere model is the systems with bounded potentials
\cite{gaussstill,bianka1,bianka2,peppino}. Since these potentials
have no singularity in the origin the behavior of such system is
strongly different from the behavior of hard spheres.

One of the most common model with bounded potential is the
Gaussian Core Model (GCM). This system is defined by the potential

\begin{equation}
  U_G(r)=\varepsilon e^{-r^2/\sigma ^2}.
\end{equation}

This potential was introduced by Stillinger \cite{gaussstill} for
simulation of the plastic crystals system . The phase diagram of
the GCM demonstrates two crystal phases - fcc and bcc
\cite{gausspd}. Starting from the densities around $\rho \sigma ^3
\approx 0.25$ the melting curve has a negative slope. It was also
shown that GCM demonstrates liquid state anomalies: density
anomaly \cite{gaussdena,gaussanom},  diffusion anomaly
\cite{gaussda,gaussanom} and structural anomaly \cite{gaussanom}.
Interestingly Stockes-Einstein relation is also violated in the
GCM system \cite{gaussseviol}.

In the article \cite{peppino} an extensive study of another model
with bounded potential was reported. This work is concerned to the
Herzian spheres system which is defined by the interparticle
potential of the form

\begin{equation}
\Phi (r)=\left\{
\begin{array}{lll}
\varepsilon (1-r/ \sigma)^{5/2} , & r\leq \sigma \\
0, & r>\sigma%
\end{array}%
\right.  \label{1}
\end{equation}

The phase diagram of the Herzian spheres system demonstrates very
complex behavior, including many crystal phases and reentrant
melting. Anomalous diffusion is also reported \cite{peppino}.

Taking into account that the behavior of the systems with negative
curvature potentials and the systems with bounded potentials is
rather different from the behavior of hard spheres a question
arises if the Rosenfeld scaling relations are applicable for such
systems.

The purpose of this article is to analyze the validity of the
entropy scaling for the systems with anomalous behavior. For our
analysis we have chosen the diffusion coefficient since it is the
simplest transport coefficient to calculate in simulation. The
article is organized as follows. In the second section we describe
the models investigated in the present work and the simulation
setup. Section III gives the results of the simulations and the
discussion of these results. Finally the section IV represents our
conclusions.

\section{II. The systems and methods}

Three systems were studied in the present work: Herzian spheres,
Gauss core model and a soft repulsive shoulder system.

For the simulation of the Herzian spheres we used a system of
$1000$ particles in a cubic simulation box. NVE MD simulation was
carried. Equations of motion were integrated by velocity Verlet
algorithm. The time step was set to $dt=0.0005$. The equilibration
run was $5 \cdot 10^5$ time steps and the production run $1.5
\cdot 10^6$ time steps. During the equilibration the velocities
were rescaled to keep the temperature constant. The diffusions
were computed via the Einstein relation for the densities from
$\rho=1.0$ till $\rho=15.0$ with the step $\Delta \rho = 0.5$.
Additional simulations were done for computing the equation of
state for the densities less then unity. Free energy of the liquid
was calculated by integrating the pressure along an isotherm
\cite{bib} and the excess entropy was obtained from the relation
$S_{ex}=\frac{U-F_{ex}}{Nk_BT}$, where $U$ is the internal
potential energy of the liquid.  The simulations were done for the
set of ten isotherms:
$T=0.01;0.02;0.03;0.05;0.1;0.15;0.2;0.25;0.3;0.5$.

In the case of GCM the system consisted of $2000$ particles. The
time step was set to $dt=0.05$. The equilibration and production
runs were $4 \cdot 10^5$ and $1 \cdot 10^6$ time steps
respectively. The diffusion was computed for the densities from
$\rho=0.1$ to $\rho=1.0$ with the step of $0.1$ for the isotherms
$T=0.04;0.05;0.06;0.07;0.1;0.2;0.3;0.4;0.5;1.0;2.0$. The
calculation of the diffusion coefficient and excess entropy was
carried in the same way as for Herzian spheres.

The last system considered in the present work is the continuous
repulsive shoulder system introduced in the article
\cite{FFGRS2008}. The potential of this system has the form

\begin{equation}
  U(r)=
  (\frac{\sigma}{r})^{14}+\frac{1}{2}\varepsilon
  \cdot[1-tanh(k_0\{r-\sigma_1\})],
\end{equation}

where $k_0=10.0$. As it was reported in the paper \cite{GFFR2009}
this system demonstrates anomalous behavior due to its quasibinary
nature. Here we extend the investigation of the diffusion anomaly
in the repulsive shoulder system with $\sigma_1 =1.35$ and check
the Rosenfeld relations for this system in the anomalous region.

For the simulation of the repulsive shoulder system we used
parallel tempering technic \cite{bib}. The details of the
simulation were described in \cite{GFFR2009}. We computed the
diffusion coefficients along different isochors starting from the
density $\rho=0.3$ to $\rho=0.8$ with the step $0.05$. A set of
$24$ temperatures between $T=0.2$ and $T=0.5$ was simulated.
Taking into account the exchange of the temperatures at the same
density more then a hundred runs at the same isochor was done.
This allowed us to collect a good statistics on the temperature
dependence of the diffusion coefficient. The diffusion coefficient
along an isochor was approximated by a $9-$th order polynome of
the temperature. The excess entropy was calculated in the way
described above.

Usually excess entropy can be well approximated by the pair
contribution only: $S_{ex}=S_{pair}+S_3+...\approx S_{pair}$,
where

\begin{equation}
 S_{pair}=-\frac{1}{2} \rho \int
 d\textbf{r}[g(\textbf{r})\ln(g(\textbf{r}))-(g(\textbf{r})-1)],
\end{equation}

where $\rho$ is the density of the system and $g(\textbf{r})$ is
the radial distribution function. We did not use the pair
contribution to the excess entropy for the GCM and Herzian spheres
because of the considerable overlap of the particles for the
bounded potentials.

Since the potentials studied in the present work have negative
curvature regions or are bounded they can not be approximated by
an one component hard spheres system. It allows us to pose a
question about the applicability of the entropy scaling to these
systems both in Rosenfeld and Dzugutov forms. Note that Dzugutov
relation (Eq. (5)) involves the size of the particles $\sigma$
which is ill defined for the negative curvature and bounded
potentials systems. This makes problematic to apply the Dzugutov
scaling rule to them. Because of this only Rosenfeld relations
were used in this work.


In this paper we use the dimensionless quantities: $\tilde{{\bf
r}}={\bf r}/ \sigma$, $\tilde{P}=P \sigma
^{3}/\varepsilon ,$ $\tilde{V}=V/N \sigma^{3}=1/\tilde{\rho},$ $\tilde{T}%
=k_{B}T/\varepsilon $. Since we use only these reduced units we
omit the tilde marks.

\section{III. Results and discussion}

This section reports the simulation results for the diffusion
coefficient and excess entropy of the three models described above
and checks the validity of the Rosenfeld relation for these
systems.

\subsection{Herzian spheres}

Low temperature behavior of the diffusion coefficient of Herzian
spheres system was already reported in the work \cite{peppino}. As
it is seen from this publication the diffusivity shows even two
anomalous regions at the temperature $T=0.01$ where diffusion
coefficient grows with growing density. In the present work the
dependence of the diffusion coefficient on density along several
isotherms was monitored. The simulation data are presented on the
Fig. 1 (a) - (b).

\begin{figure}
\includegraphics[width=8cm, height=8cm]{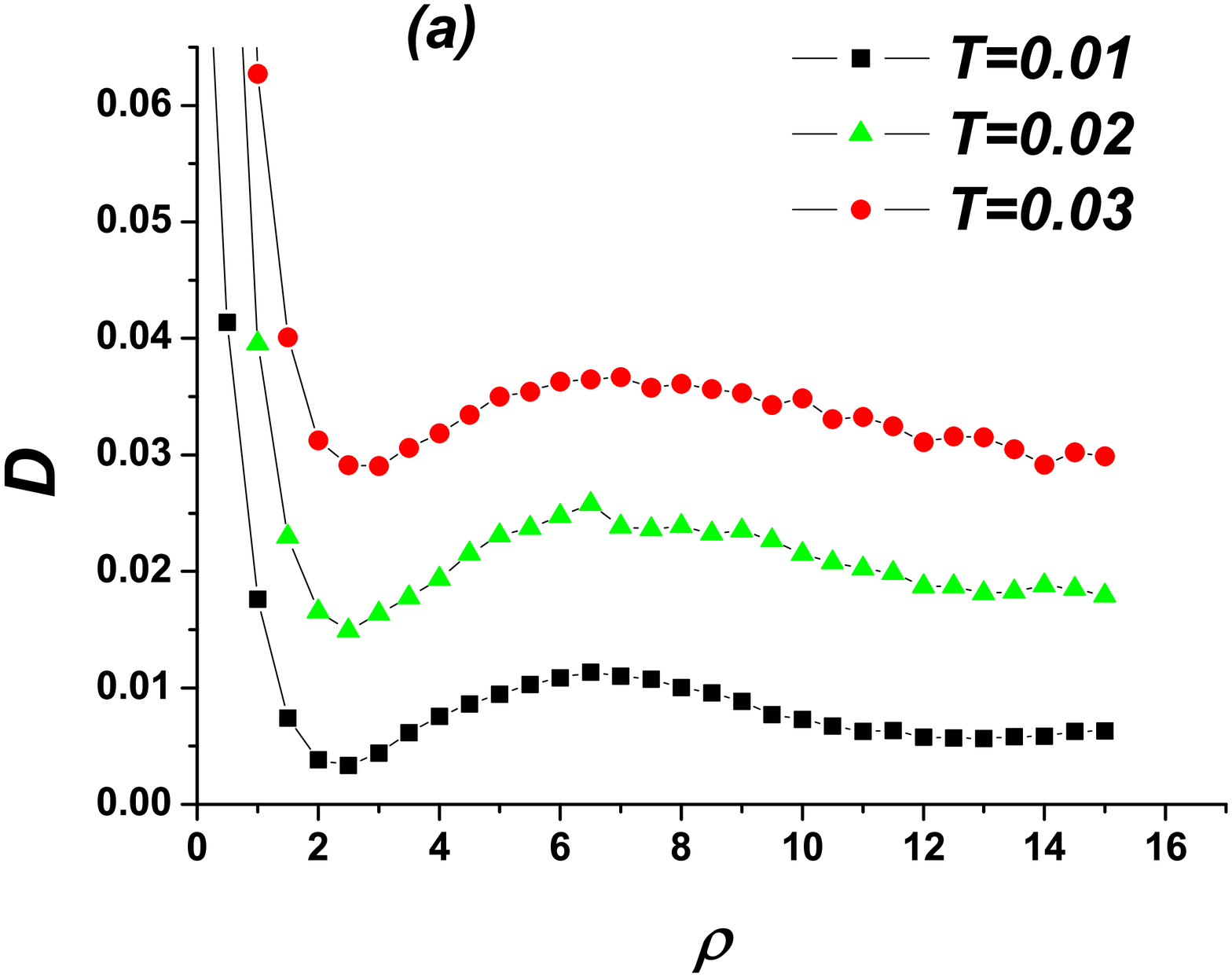}%

\includegraphics[width=8cm, height=8cm]{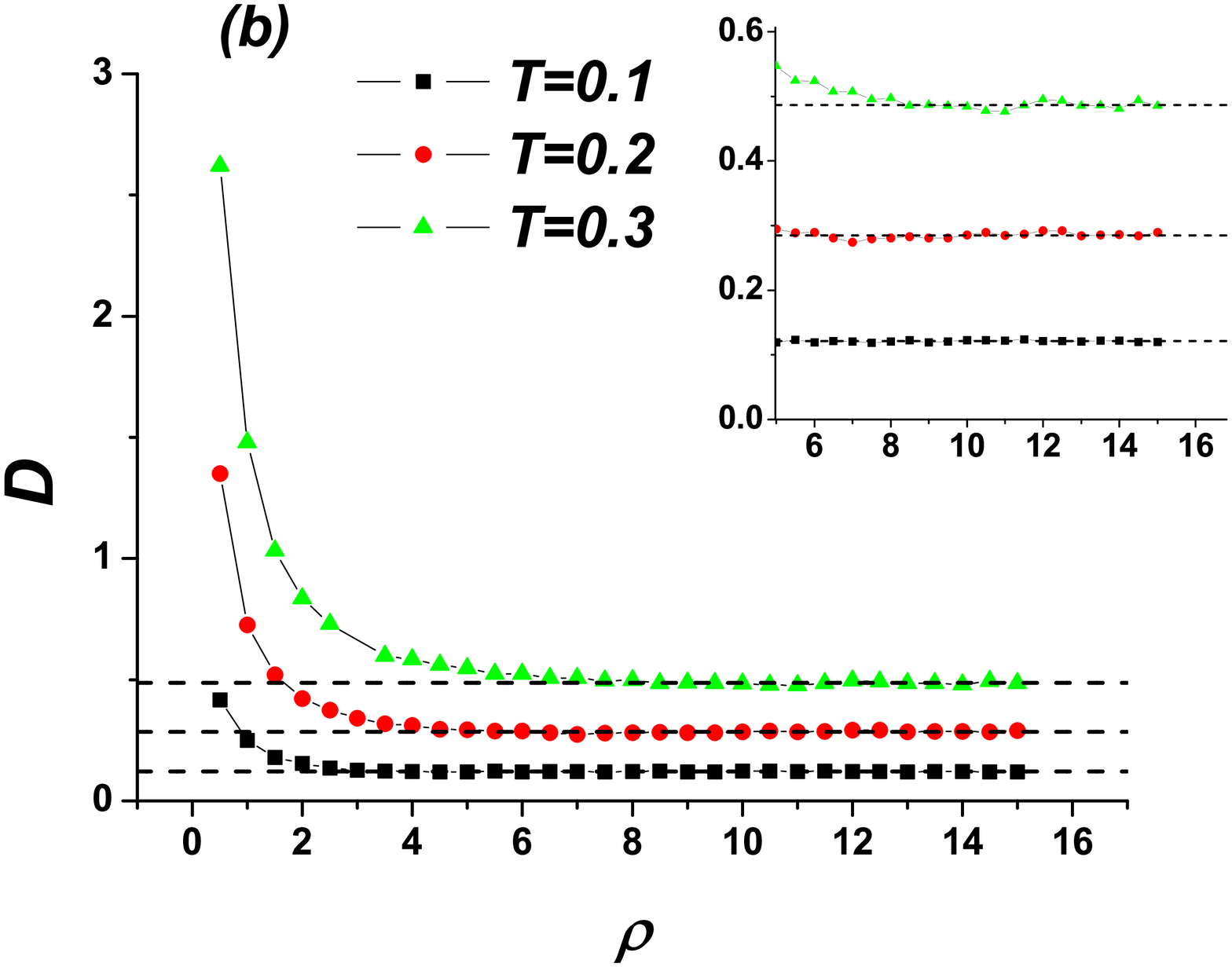}%
\caption{\label{fig:fig1}  Diffusion coefficient of Herzian
spheres for a set of isotherms. (a) - $T=0.01;0.02$ and $0.03$;
(b) - $T=0.1;0.2$ and $0.3$. The inset of the figure (b) showes
the high density behavior of diffusion coefficient.}
\end{figure}

One can see from these figures that at low temperatures (Fig.
1(a)) the diffusion is non monotonic, while at high temperatures
it monotonically decays with increasing density (Fig. 1(b)) and
comes to a constant value (see inset of the Fig. 1(b)).

It is worth to note that the melting temperatures of Herzian
spheres reported in the work \cite{peppino} are of the order of
$10^{-3}$, so the temperatures about $0.1$ are extremely high for
this model. This is easily seen from the Fig. 2 (a) - (b) where
the radial distribution functions for the density $\rho=6.0$ are
shown for the same set of temperatures. One can see that at
$T=0.01$ the liquid has short range structure which rapidly decays
with increasing temperature. At the temperature $T=0.1$ the liquid
looks almost like an ideal gas since $g(r)$ comes to unity very
quickly.

\begin{figure}
\includegraphics[width=8cm, height=8cm]{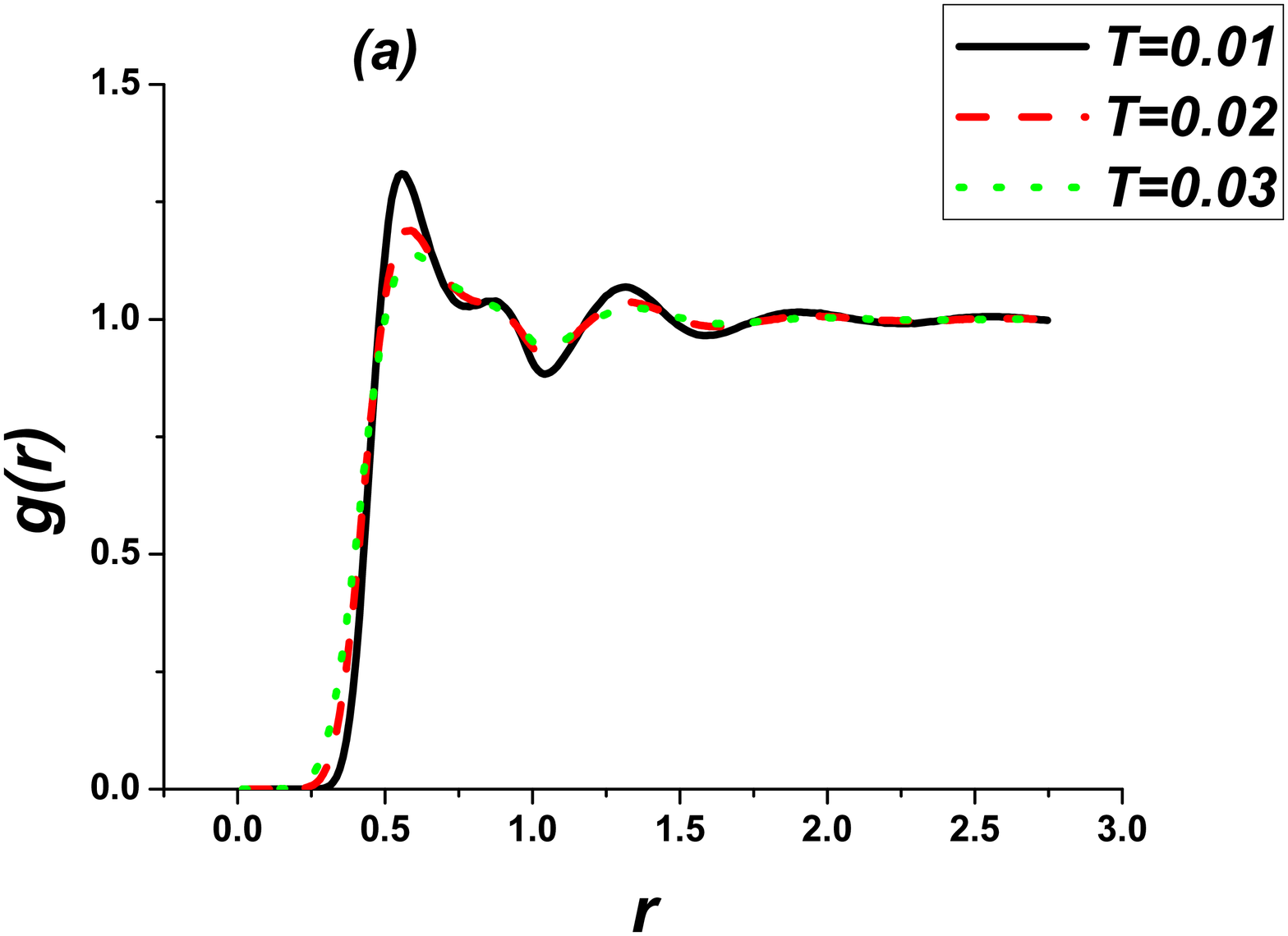}%

\includegraphics[width=8cm, height=8cm]{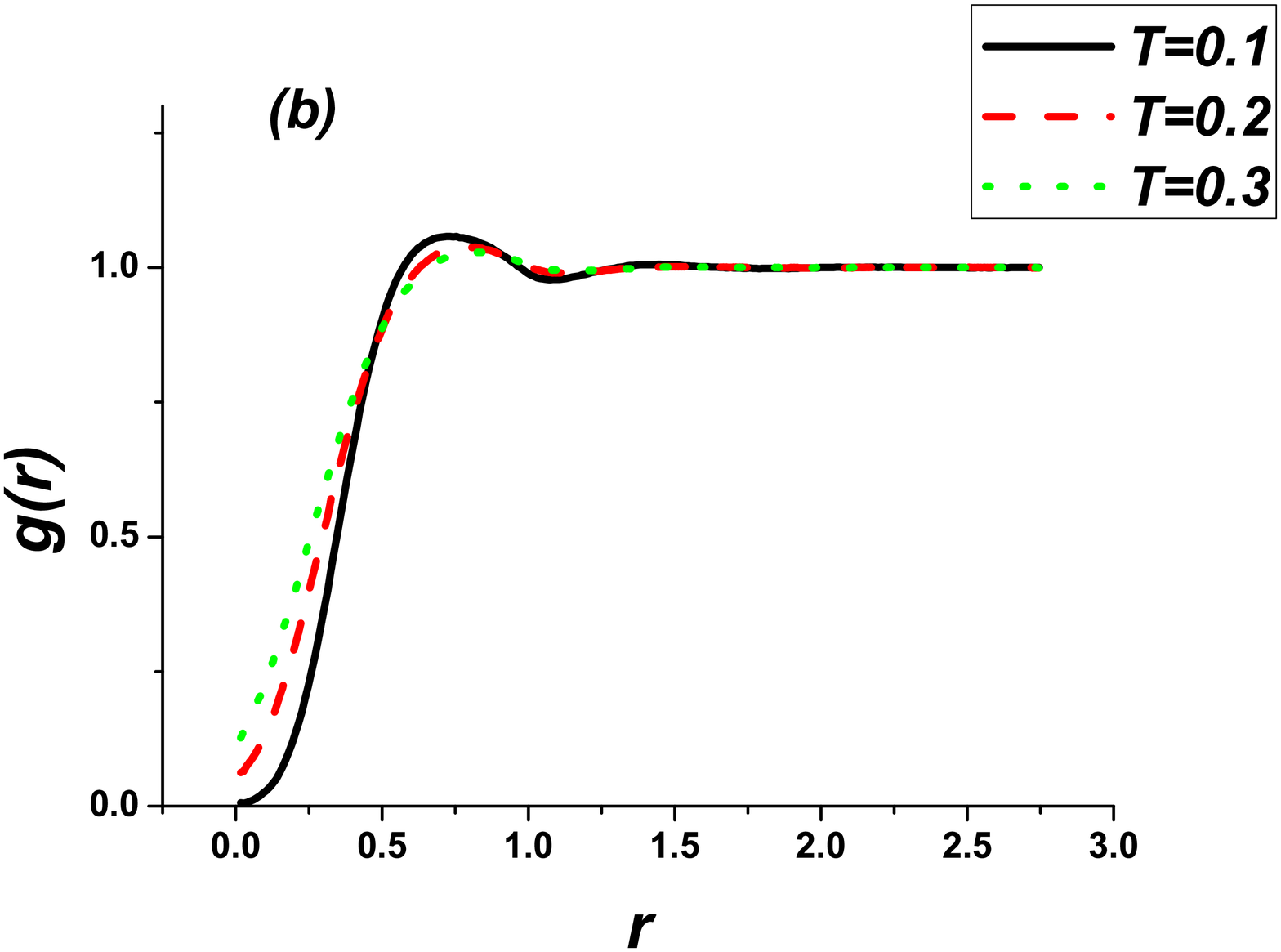}%
\caption{\label{fig:fig2}  Radial distribution functions of
Herzian spheres at $\rho=6.0$ and a set of temperatures. (a) -
$T=0.01;0.02$ and $0.03$; (b) - $T=0.1;0.2$ and $0.3$.}
\end{figure}

Excess entropy also shows non monotonic dependence on density
along an isotherm (Fig. 3(a) -(b)). One can see from these figures
that at low temperatures excess entropy has two minima and a
maxima in the investigated density range while at high temperature
the first minima is depressed and the curves just change the slope
smoothly.

\begin{figure}
\includegraphics[width=8cm, height=8cm]{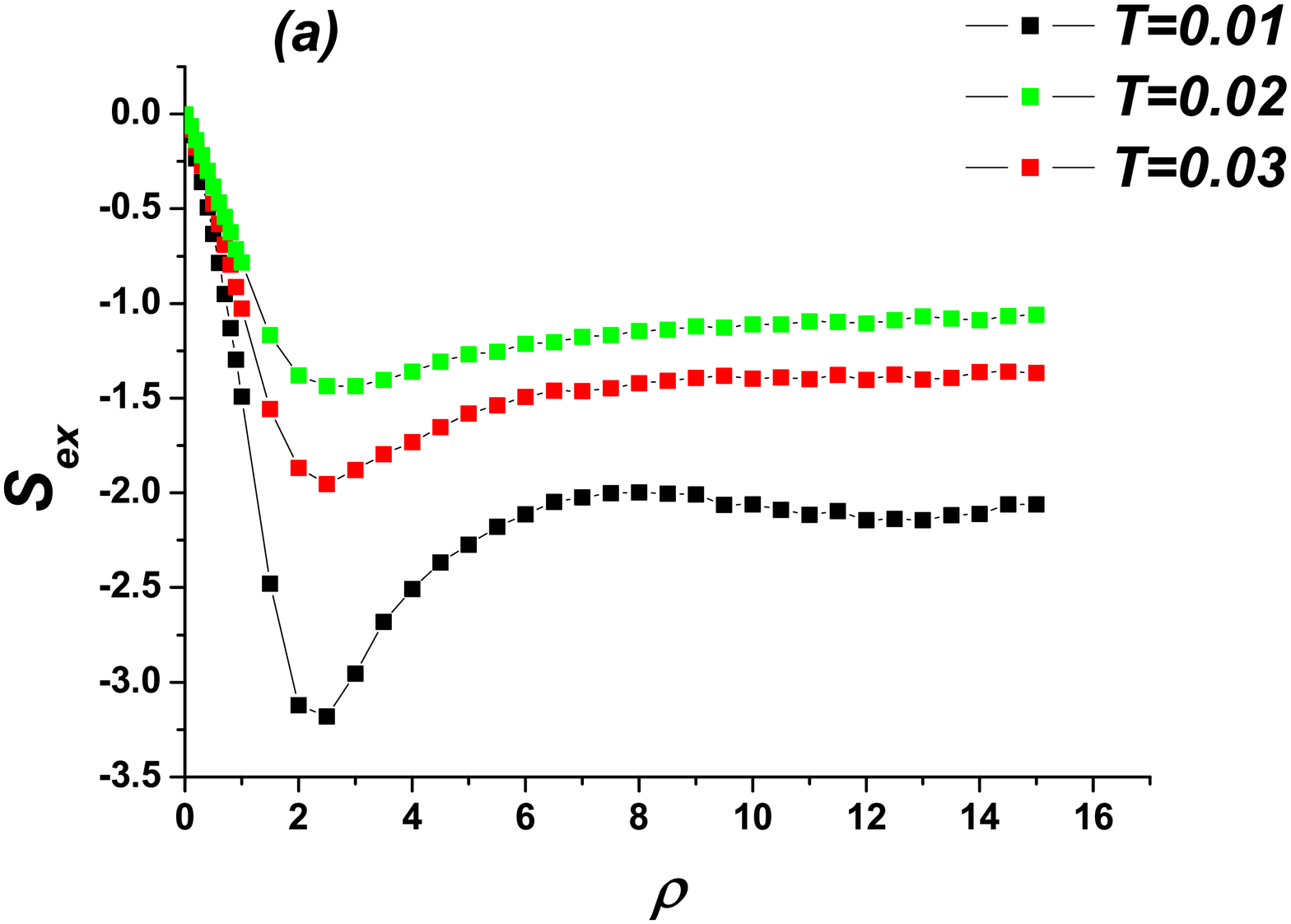}%

\includegraphics[width=8cm, height=8cm]{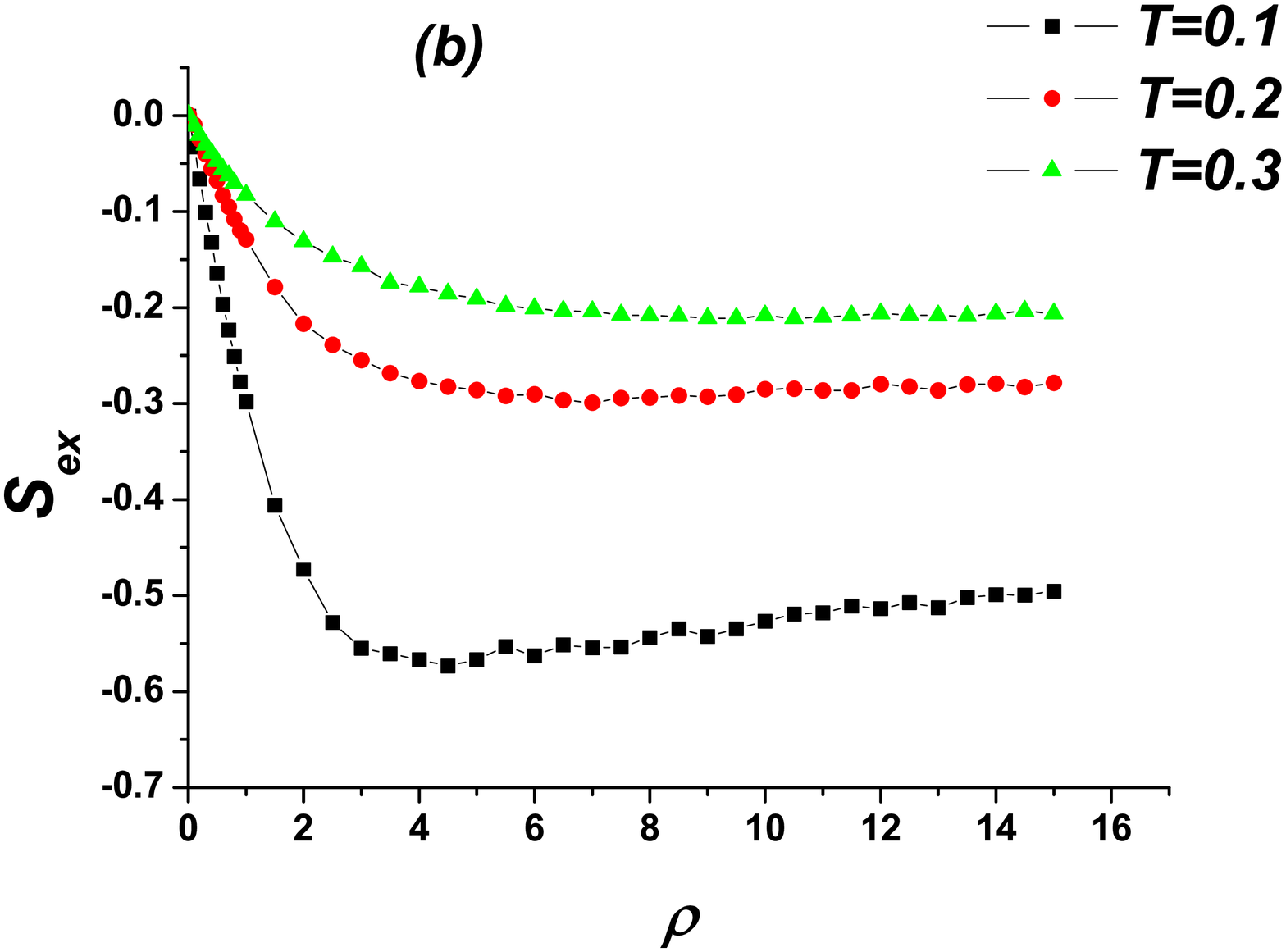}%
\caption{\label{fig:fig3}  Excess entropy along and a set of
isotherms. (a) - $T=0.01;0.02$ and $0.03$; (b) - $T=0.1;0.2$ and
$0.3$.}
\end{figure}

Now we turn to the Rosenfeld relation for the Herzian spheres
system. The dependence of the reduced diffusion (see formula (1))
on the excess entropy along some isotherms is shown in the Fig. 4
(a) - (b). Looking at the curve for $T=0.01$ (Fig. 4 (a)) one can
divide it into three distinct regions with different slopes which
we denote as '1', '2' and '3'. The density increase corresponds to
moving along the curves from right to left, i.e. region 2
corresponds to the higher densities then 1, and 3 - higher
densities then 2. As is seen from the plots the region 3 rapidly
disappears with increasing the temperature. Already at $T=0.03$
this region is negligibly small. Recall from the figures 1 and 3
that at low temperatures both diffusion and excess entropy behave
non monotonically while at growing the temperature this effect
disappears. This leads to the depression of the region 3 in the
Fig. 4 (a).

\begin{figure}
\includegraphics[width=8cm, height=8cm]{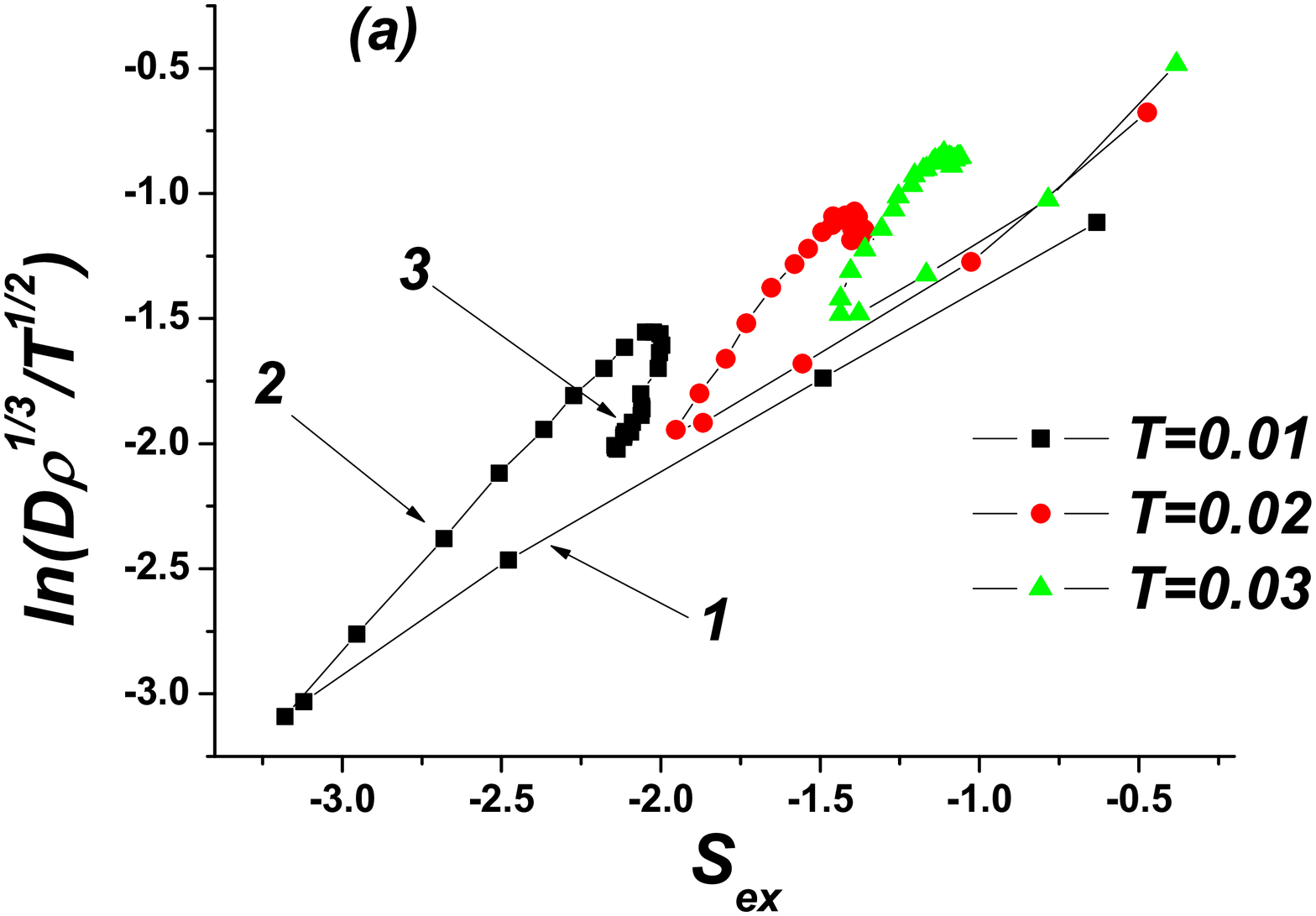}%

\includegraphics[width=8cm, height=8cm]{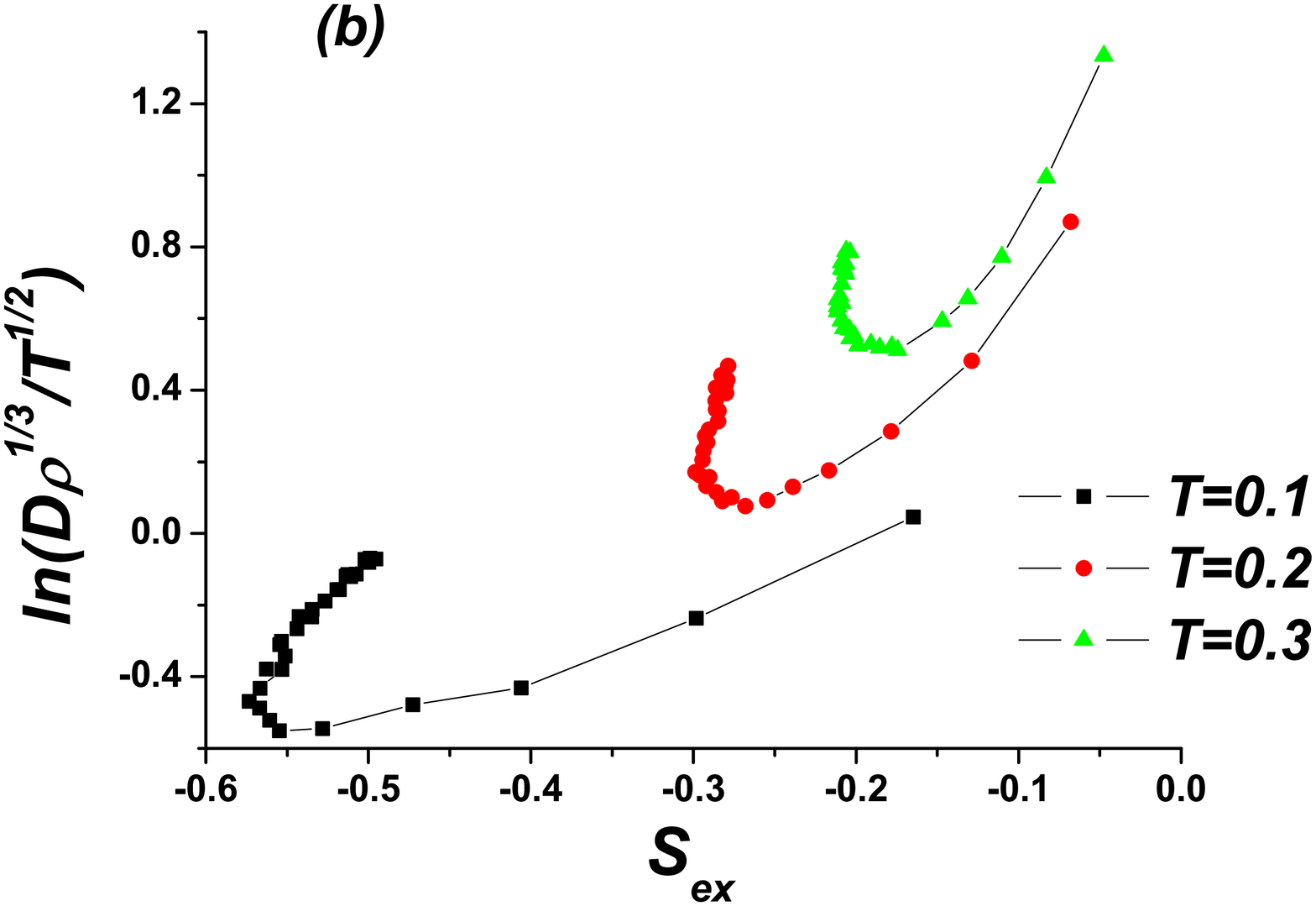}%
\caption{\label{fig:fig4}  The dependence of reduced diffusion on
the excess entropy (see formulas (1) and (4)) along some isotherms
(a) - $T=0.01;0.02$ and $0.03$; (b) - $T=0.1;0.2$ and $0.3$.}
\end{figure}

The region 2 is rather stable. As one can see from the Fig. 4 (b)
this region also becomes less developed with increasing the
temperature, but it still preserves even for high temperatures. It
makes the excess entropy scaling curve consisting from two parts
of different slope and a cross region.

\begin{figure}
\includegraphics[width=8cm, height=8cm]{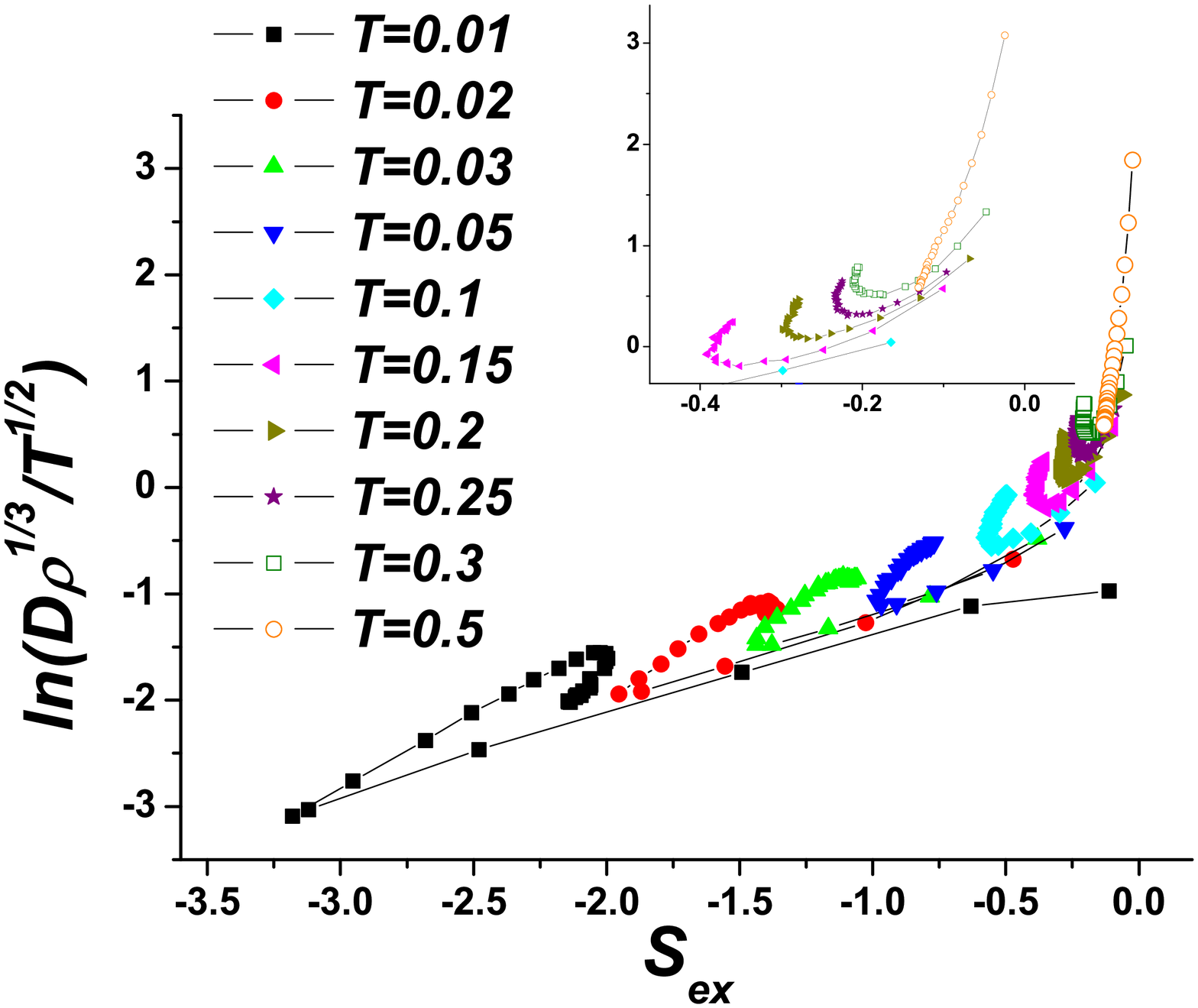}%

\caption{\label{fig:fig5} Rosenfeld excess entropy scaling of
diffusion coefficient for a set of ten isotherms for Herzian
spheres system. The inset shows several high temperature isotherms
in the enlarged scale.}
\end{figure}

Fig. 5 summarizes all the results obtained for Herzian spheres
system. Ten different isotherms are shown there. As one can see
only at the temperature as high as $0.5$ the Rosenfeld linear
relation between the logarithm of the reduced density and the
excess entropy becomes valid. Remember that the melting
temperature is of the order of $0.005$, that is $100$ times
smaller. It allows to say that the Rosenfeld relation for
diffusion coefficient of Herzian spheres is valid only in the
infinitely high temperature limit.

\subsection{ Gaussian Core Model}

The results for the GCM are qualitatively similar to the case of
Herzian spheres. Because of this we do not explain them in detail.
Fig. 6(a) and (b) shows the diffusion coefficient for the GCM
system for a set of six isotherms. One can see that for low
temperatures starting from the densities approximately $0.3$ the
diffusion coefficient demonstrate anomalous growth with increase
of the density. We expect that at higher densities the curve bends
downward but in the present study we have not measured so high
densities for this model. At high temperatures the diffusion
monotonically decreases with increasing the density.

\begin{figure}
\includegraphics[width=8cm, height=8cm]{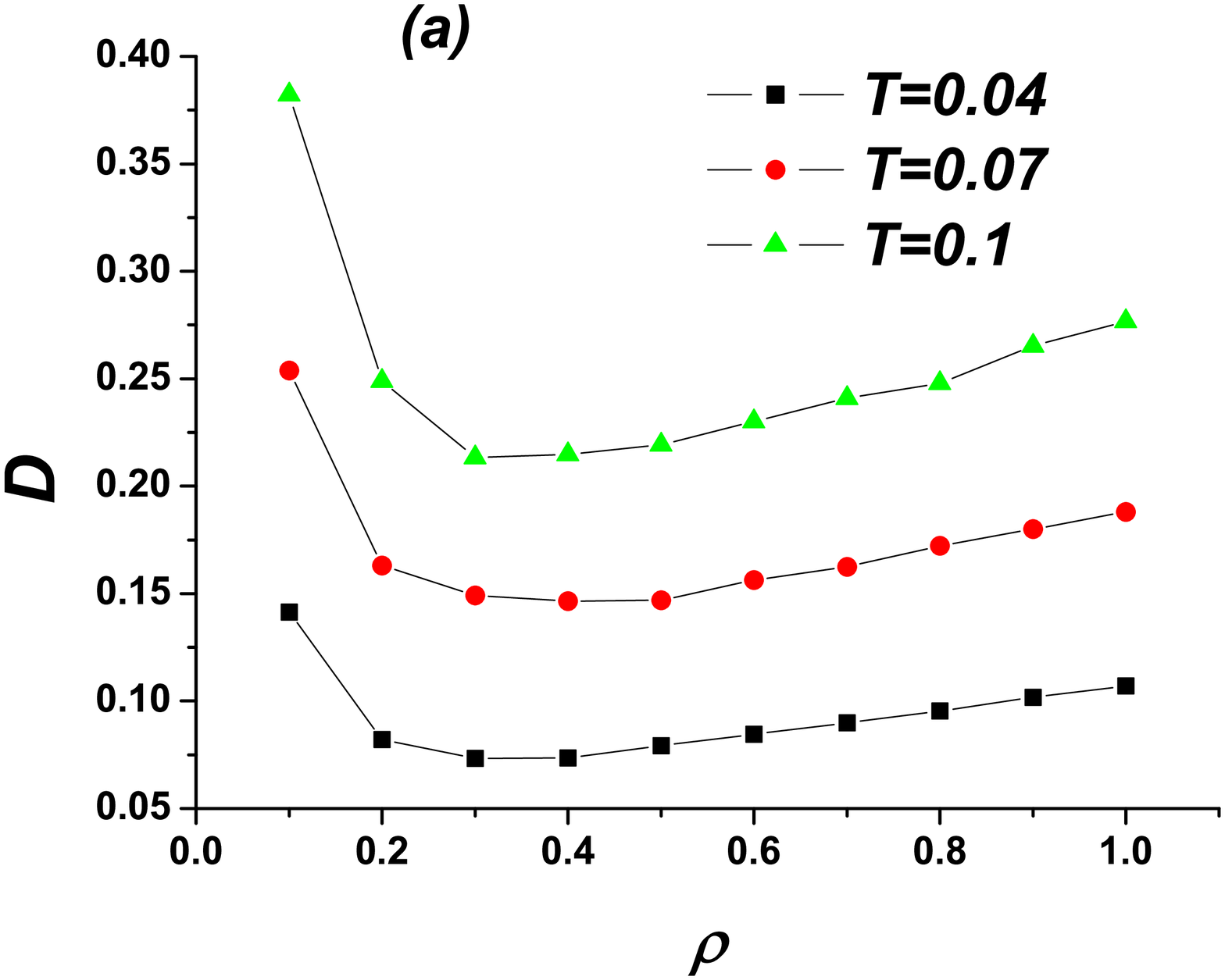}%

\includegraphics[width=8cm, height=8cm]{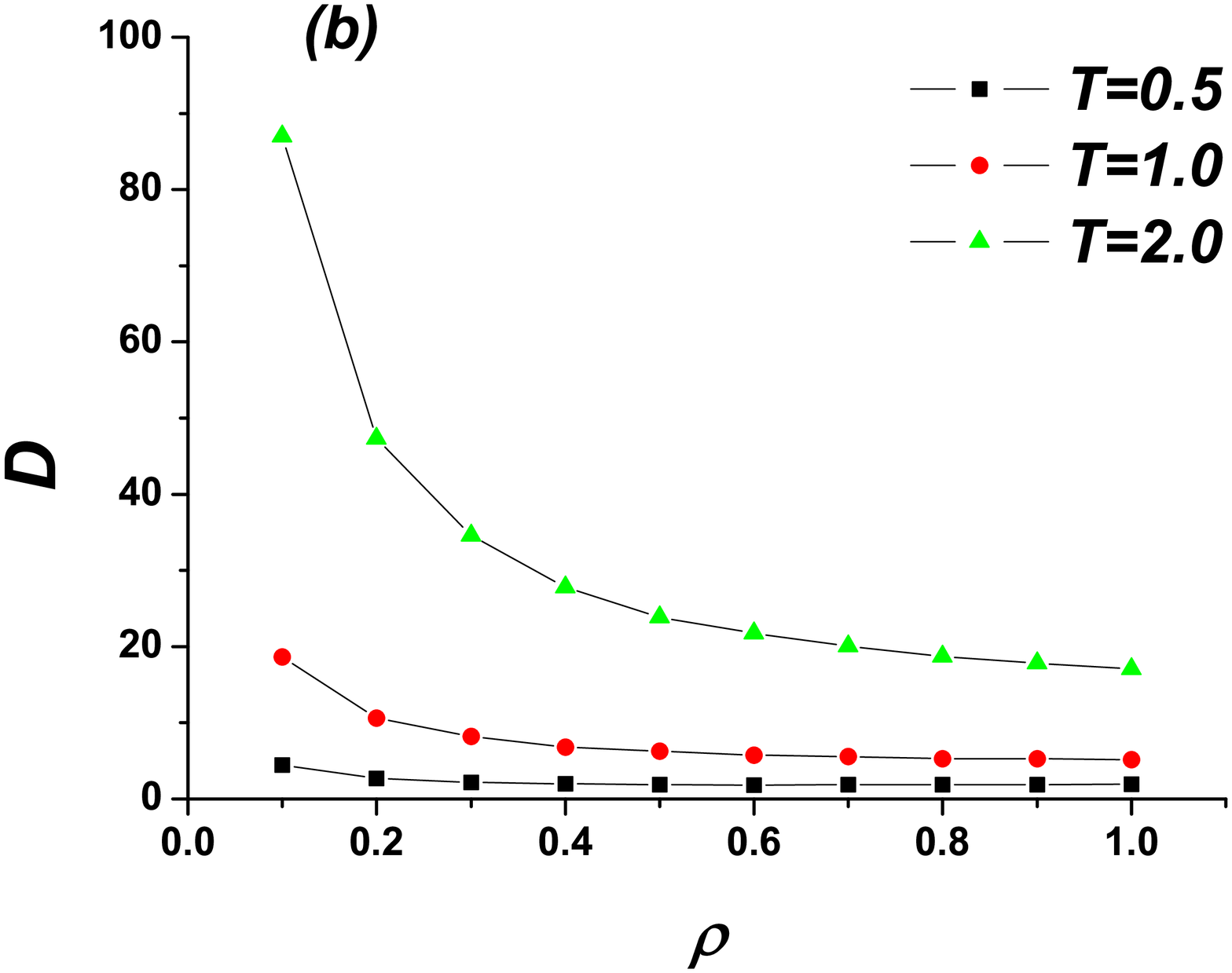}%
\caption{\label{fig:fig6}  Diffusion coefficient of GCM for a set
of isotherms. (a) - $T=0.04;0.07$ and $0.1$; (b) - $T=0.5;1.0$ and
$2.0$.}
\end{figure}

The structure of the liquid rapidly decays with the temperature
increase. This is shown in the Fig. 7(a) - (b). One can see that
at the temperature $T=1.0$ $g(r)$ is equal to unity almost in the
whole range of the distances $r$.

\begin{figure}
\includegraphics[width=8cm, height=8cm]{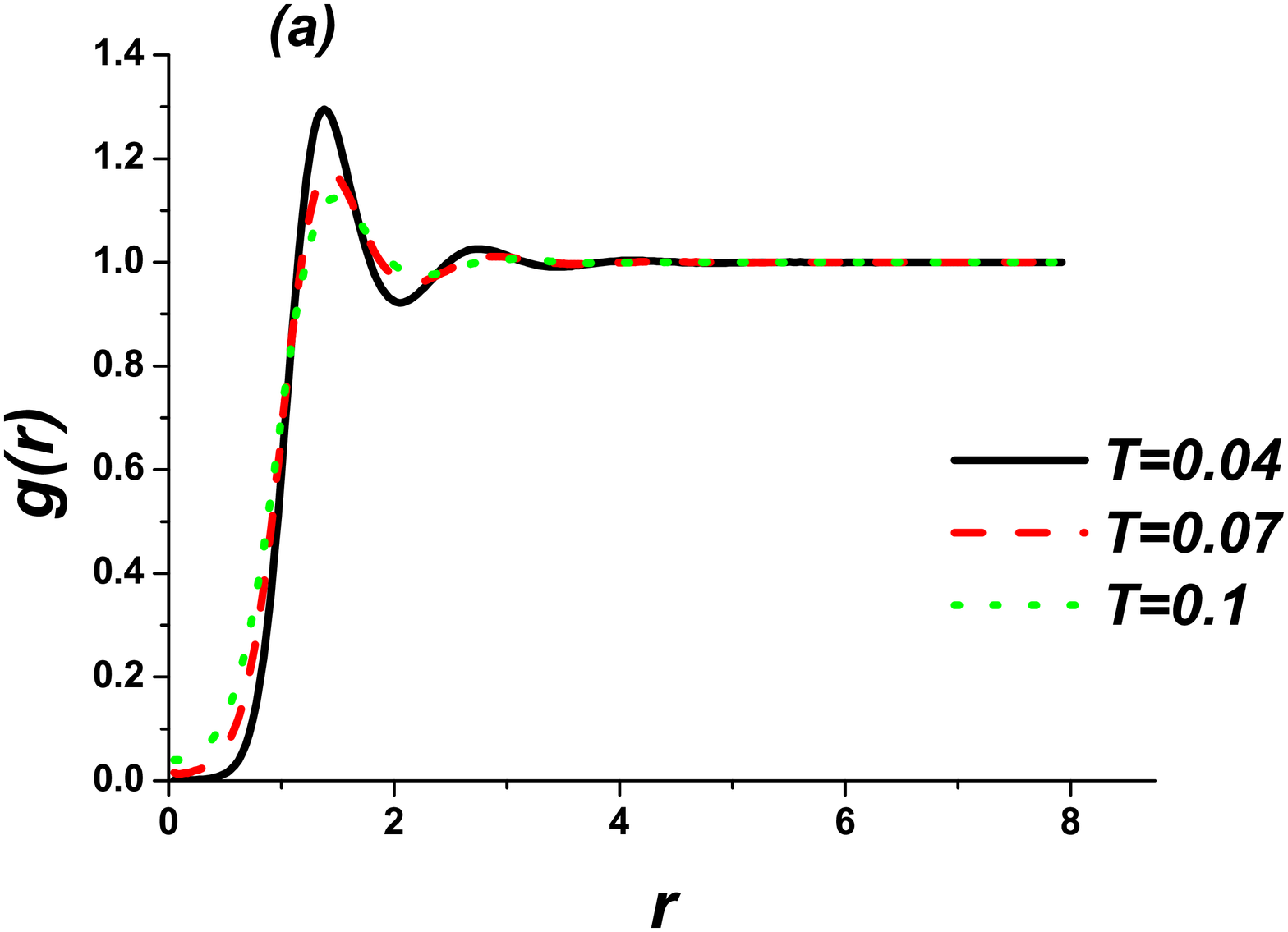}%

\includegraphics[width=8cm, height=8cm]{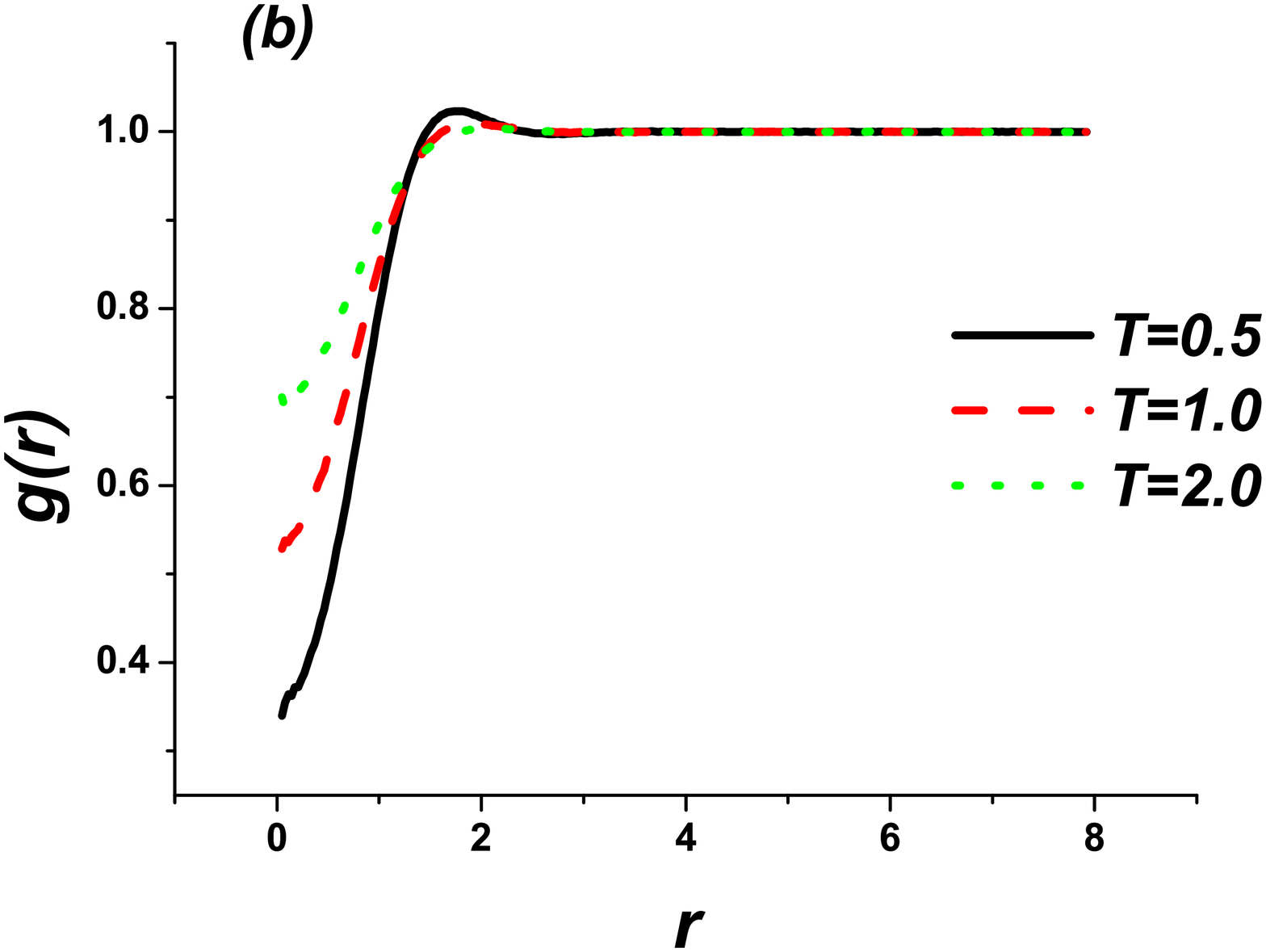}%
\caption{\label{fig:fig7}  Radial distribution functions of GCM
for a set of isotherms at the density $\rho=0.5$ (a) -
$T=0.04;0.07$ and $0.1$; (b) - $T=0.5;1.0$ and $2.0$.}
\end{figure}

Like the diffusion coefficient the excess entropy has a minimum at
the low temperatures, but with increasing the temperature it
becomes monotonically decreasing function of the density (Fig.
8(a) - (b)).

\begin{figure}
\includegraphics[width=8cm, height=8cm]{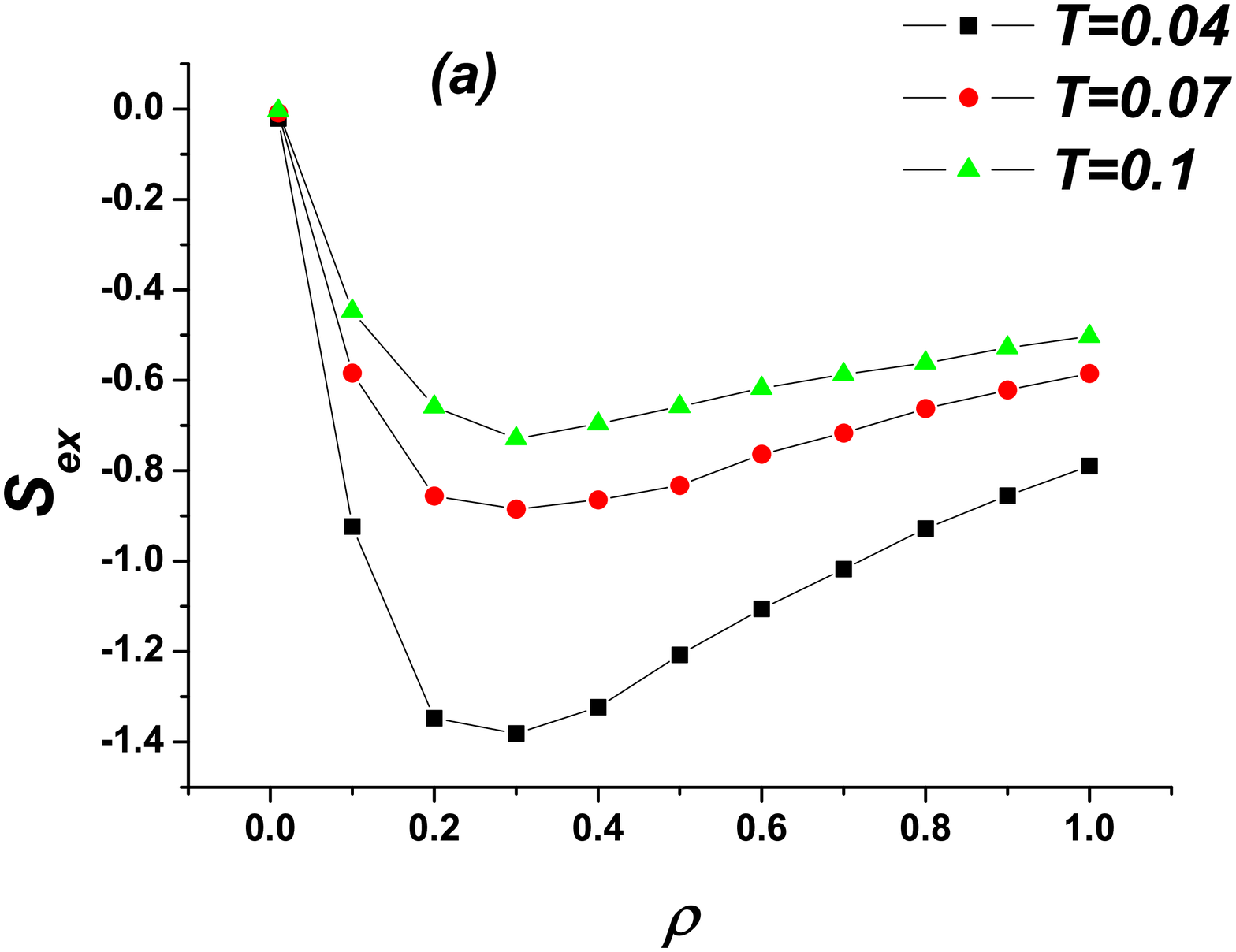}%

\includegraphics[width=8cm, height=8cm]{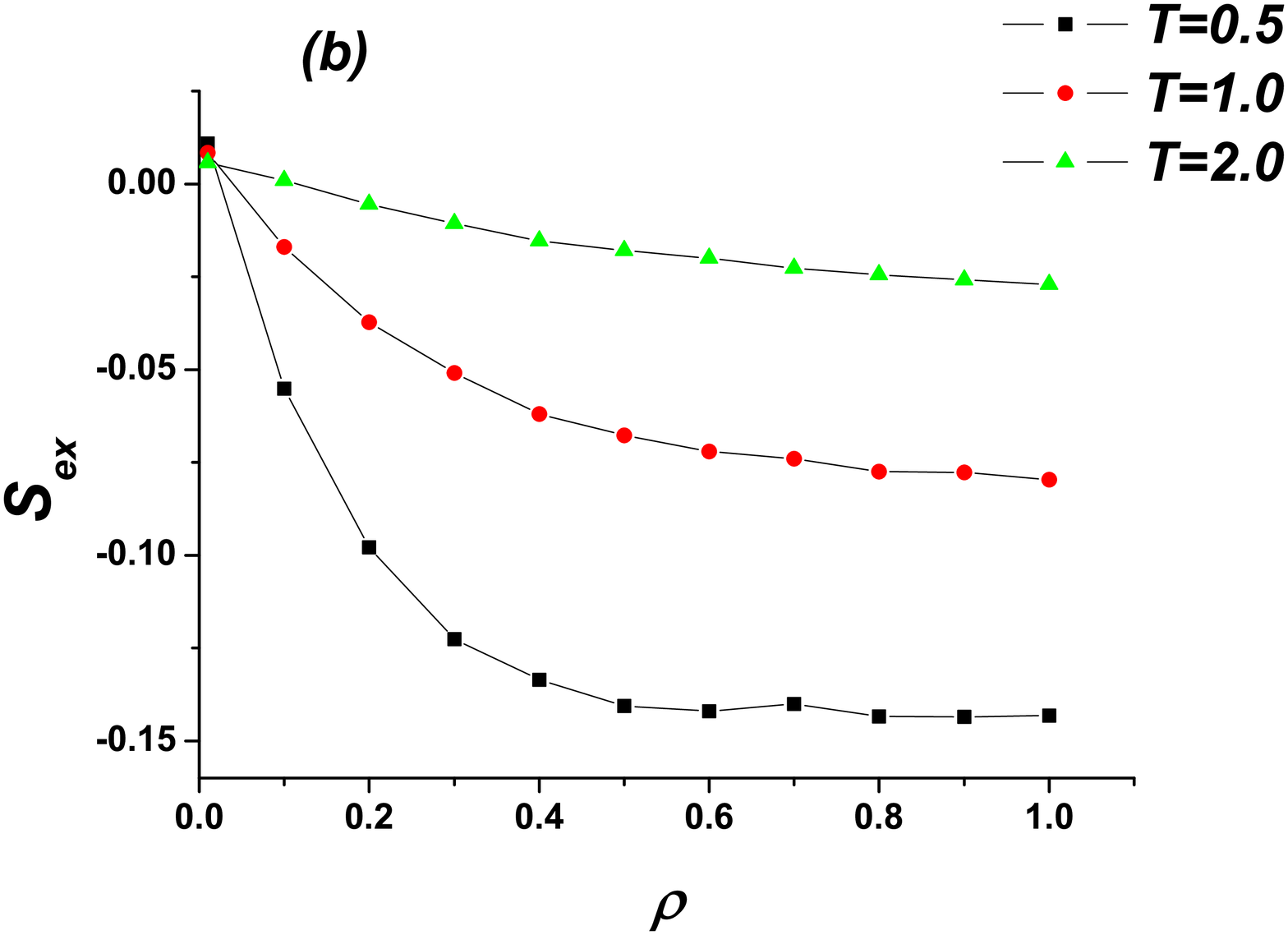}%
\caption{\label{fig:fig8}  Excess entropy of GCM for a set of
isotherms. (a) - $T=0.04;0.07$ and $0.1$; (b) - $T=0.5;1.0$ and
$2.0$.}
\end{figure}

Fig. 9(a) - (b) demonstrates the Rosenfeld relation for the GCM
for the same set of isotherms. Comparing to the case of Herzian
spheres there is no region '3' in the low temperature curves of
the GCM. We suppose that this region corresponds to the higher
densities which we do not consider in the present study. At low
temperatures most of the points belong to the region '2'. However
this region depresses with the temperature increase. Although even
at the temperature as high as $T=2.0$ which is around $200$ times
higher then melting temperature of GCM the curve still
demonstrates a bend from a straight line at high densities.

\begin{figure}
\includegraphics[width=8cm, height=8cm]{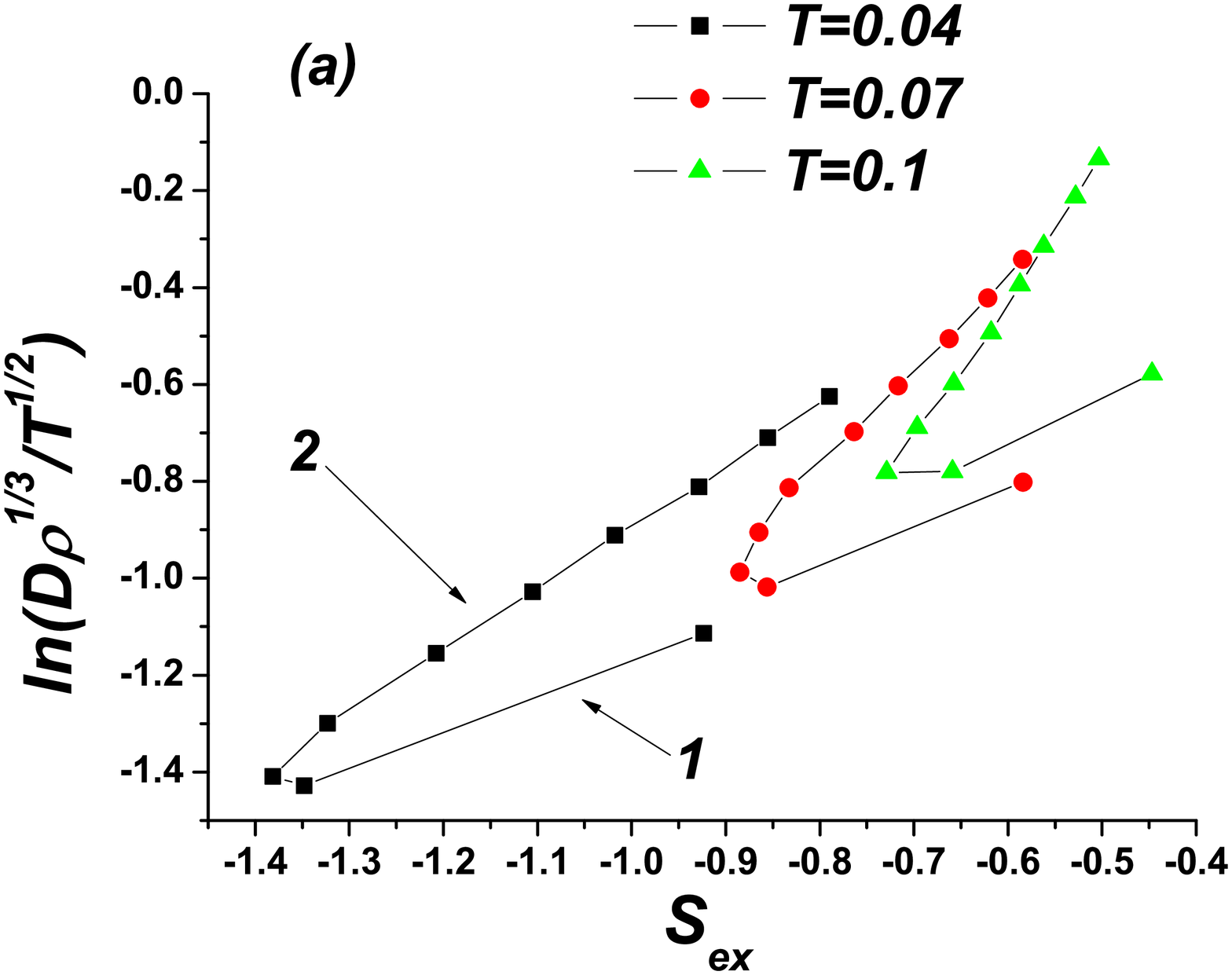}%

\includegraphics[width=8cm, height=8cm]{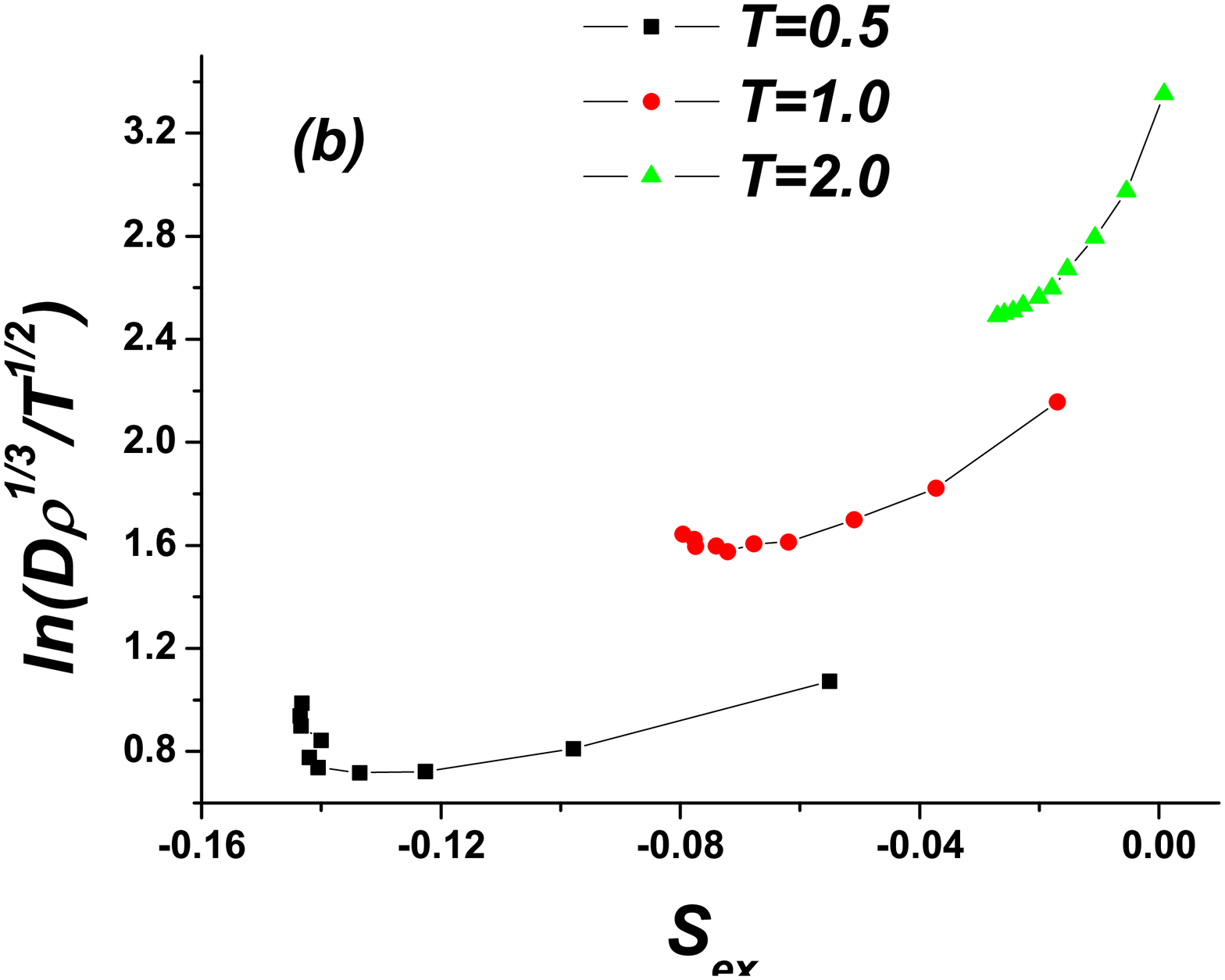}%
\caption{\label{fig:fig9}  Excess entropy of GCM for a set of
isotherms. (a) - $T=0.04;0.07$ and $0.1$; (b) - $T=0.5;1.0$ and
$2.0$.}
\end{figure}

Finally Fig. 10 shows the whole set of isotherms investigated in
the present study. As one can see from this figure the system
comes more close to the Rosenfeld relation with increasing the
temperature. One can expect that at infinitely high temperatures
the excess entropy relation is valid for the GCM, however in the
temperatures range studied here, i.e. to approximately $200 \cdot
T_{mels}$ a deviation from the linear behavior is still observed.

\begin{figure}
\includegraphics[width=8cm, height=8cm]{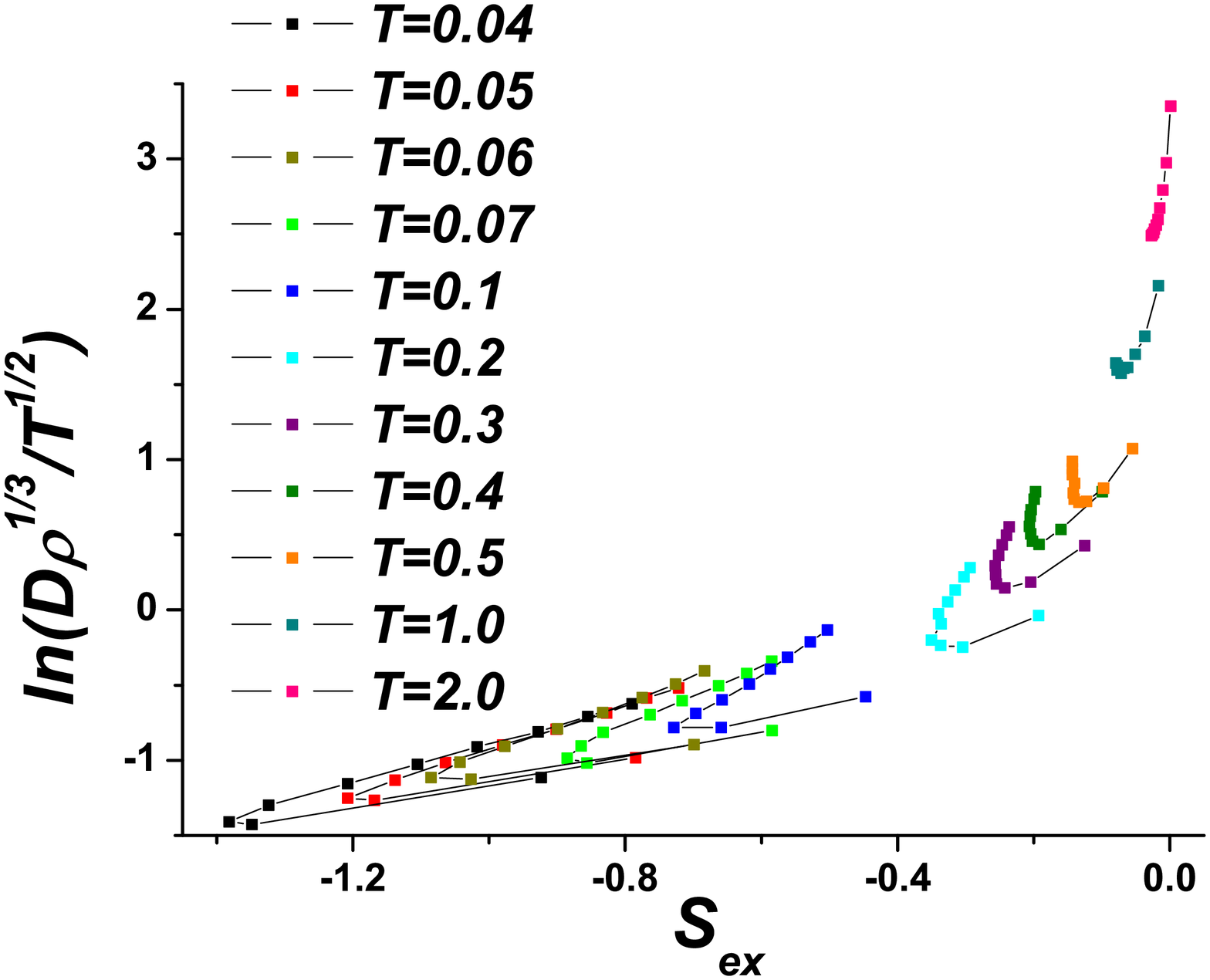}%

\caption{\label{fig:fig10}  Reduced diffusion coefficient (formula
(1)) of GCM for a set of eleven isotherms}
\end{figure}

\subsection{ Soft Repulsive Shoulder Model}

The last model considered in the present work is the soft
repulsive shoulder potential (8). In the works
\cite{FFGRS2008,GFFR2009} it has already been shown that this
system demonstrates anomalous behavior of diffusion at low
temperatures and densities corresponding to the region where a
competition between the characteristic length scales $\sigma$ and
$\sigma_1$ takes place. This competition gives the great
complexity of the phase diagram of the system
\cite{FFGRS2008,GFFR2009}, so one can expect that the
thermodynamic quantities, in particular entropy which is of the
interest of the present study, also have a complex behavior in
this region of densities. Taking into account the complex behavior
of both entropy and diffusion coefficient it is interesting to
check the Rosenfeld relation for this system.

Fig. 11 represents the diffusion coefficient for the soft
repulsive shoulder system with $\sigma_1=1.35$ for a set of
temperatures and densities. One can see that at $T=0.25$ an
inflection point in the diffusion coefficient curve occurs which
then develops into a loop ($T=0.2$).

\begin{figure}
\includegraphics[width=8cm, height=8cm]{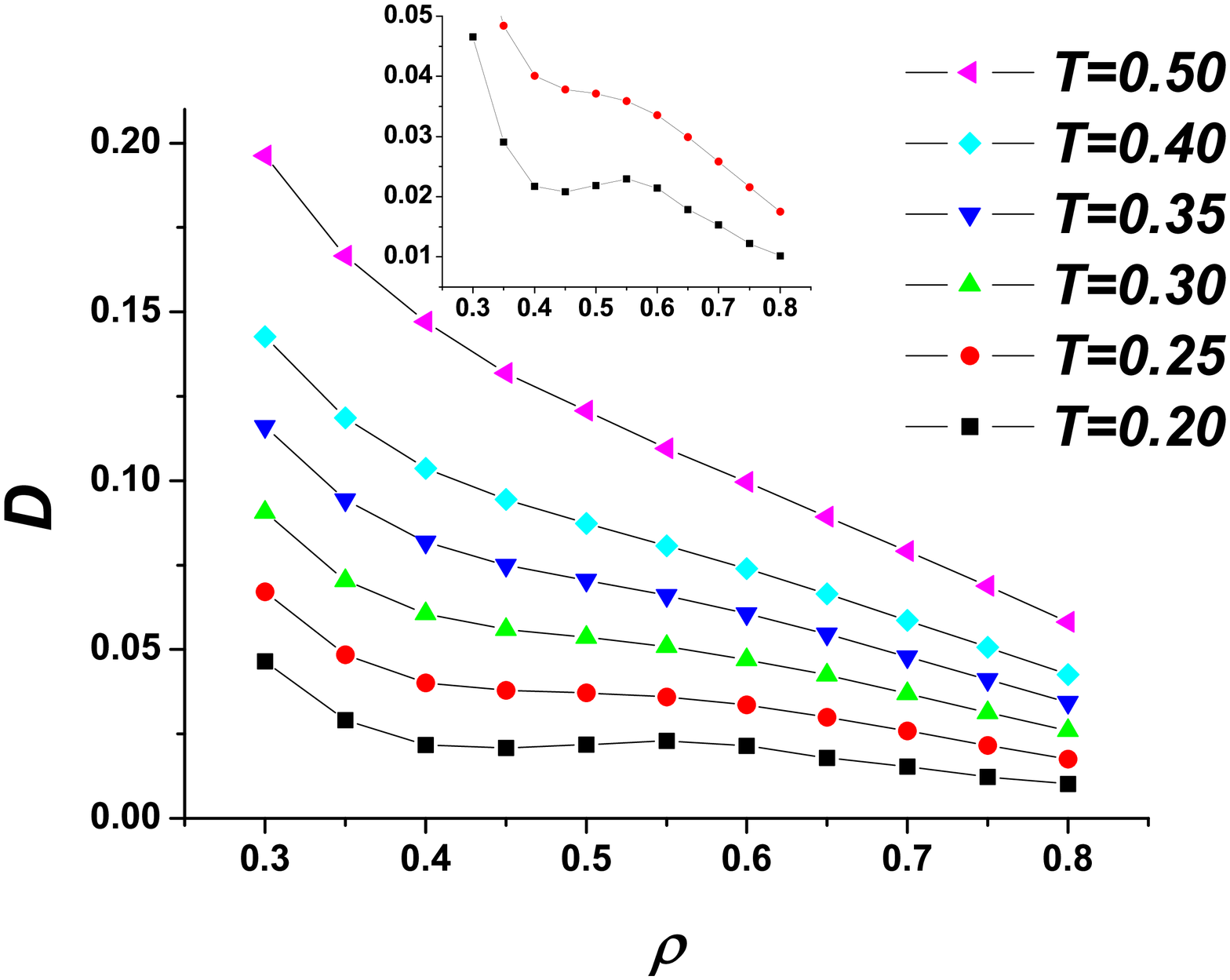}%

\caption{\label{fig:fig11} Diffusion coefficient for soft
repulsive shoulder system with $\sigma_1=1.35$ along several
isotherms. The inset enlarges the isotherms $T=0.5$ (squares) and
$T=0.25$ (circles).}
\end{figure}

Both pair and full excess entropies were considered for this
potential. Fig. 12 (a) and (b) show the behavior of the entropies
along two isotherms. One can see that both at high and low
temperature the difference between excess entropy and pair
contribution to it is rather large. This discrepancy is small at
low densities, but greatly increases at the density about $0.4$.
Note that this density corresponds to a character distance $l\sim
1/\rho^{1/3} \simeq 1.35$, that is $l\simeq \sigma_1$. It allows
to conclude that the interplay of the distances starts at this
density and it is this interplay which makes the excess entropy
and pair excess entropy difference to increase rapidly.

\begin{figure}
\includegraphics[width=8cm, height=8cm]{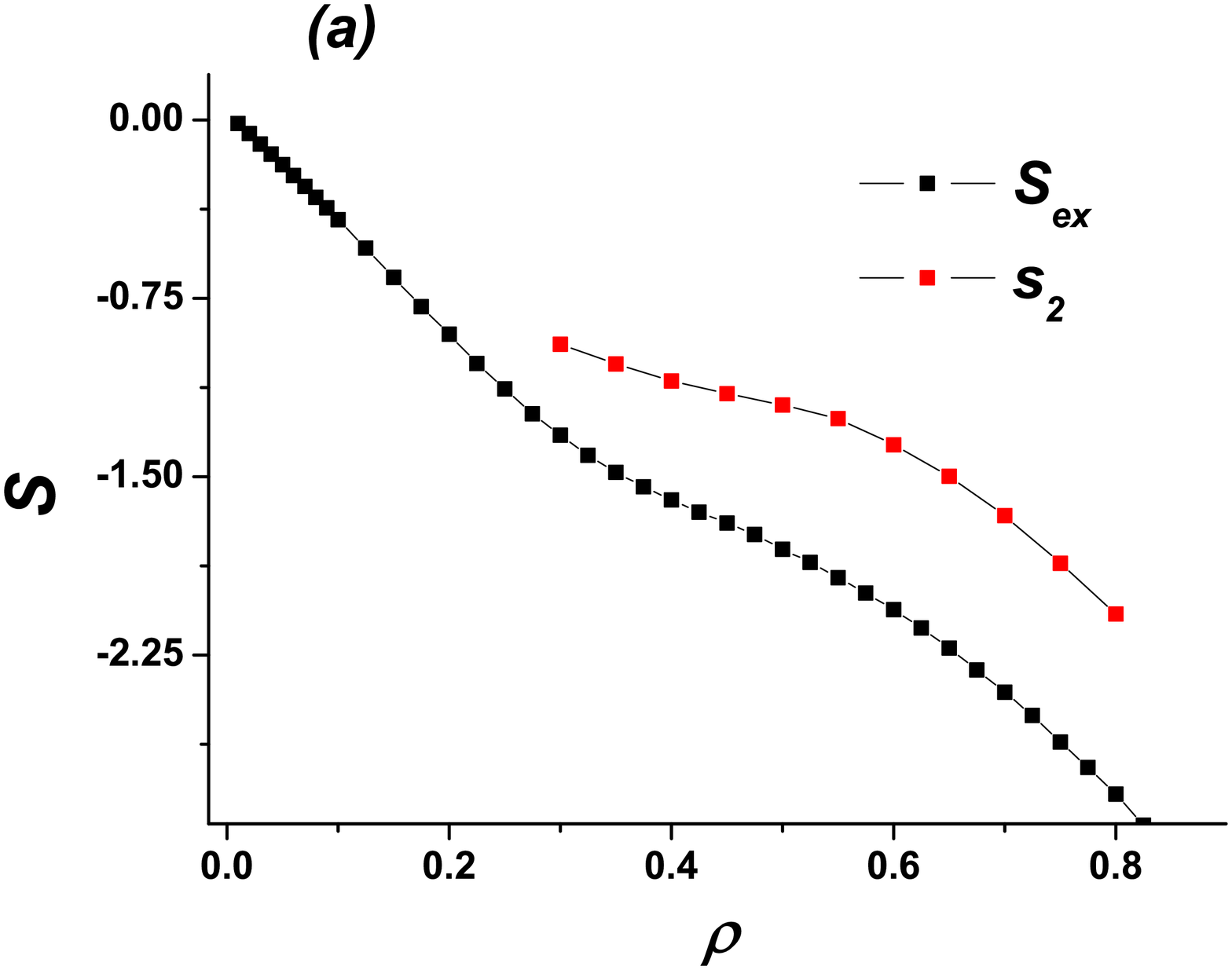}%

\includegraphics[width=8cm, height=8cm]{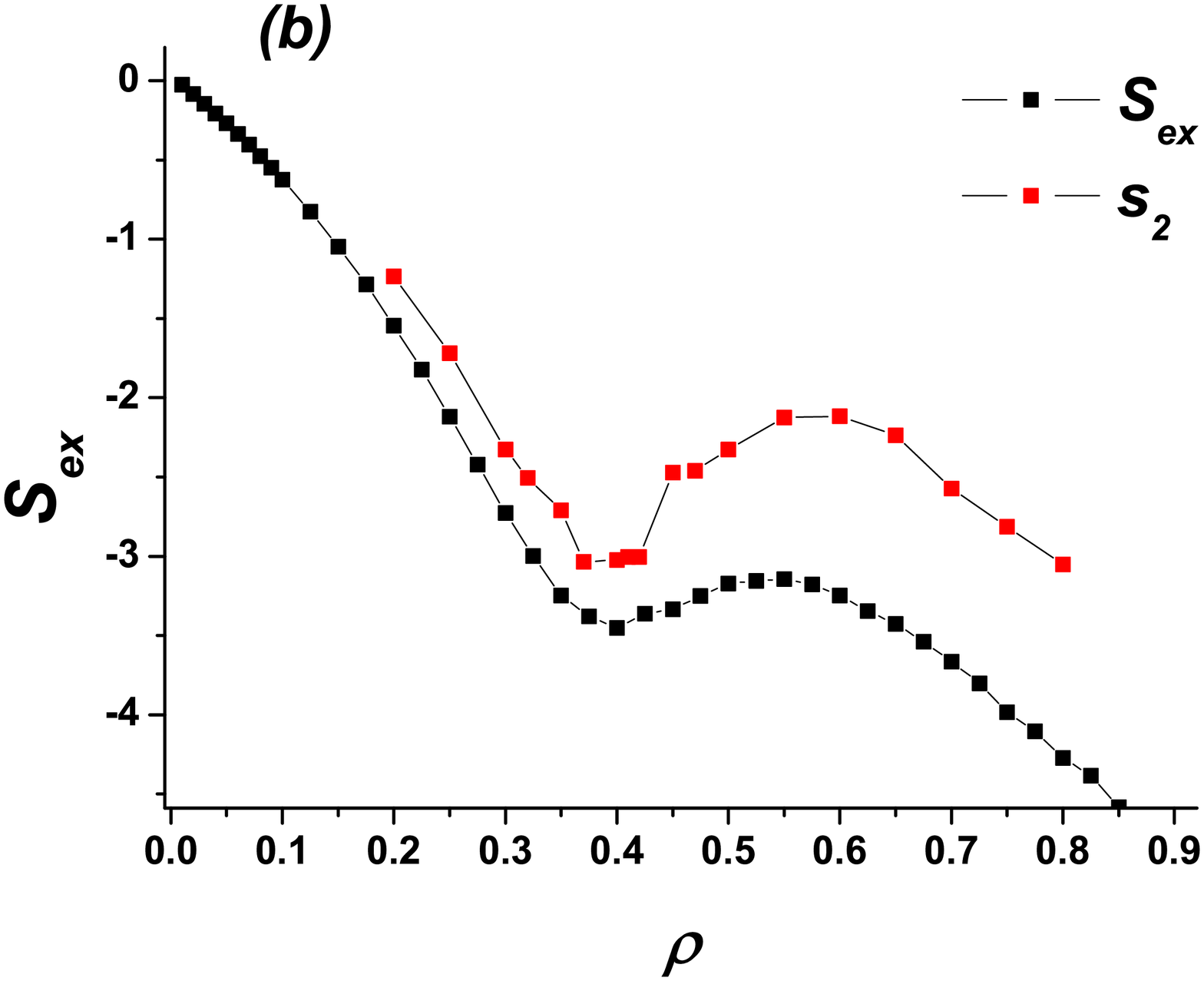}%
\caption{\label{fig:fig12} Excess entropy and pair excess entropy
for the soft repulsive shoulder system for (a) $T=0.5$ and (b)
$T=0.2$.}
\end{figure}

Fig. 13 shows the diffusion coefficient scaling with the pair part
of the excess entropy. As is seen from the figures even at high
temperatures the curve is not straight while at low temperatures
the curve becomes very strange. Definitely the exponential
relation between the diffusion coefficient and pair excess entropy
is not valid.

\begin{figure}
\includegraphics[width=8cm, height=8cm]{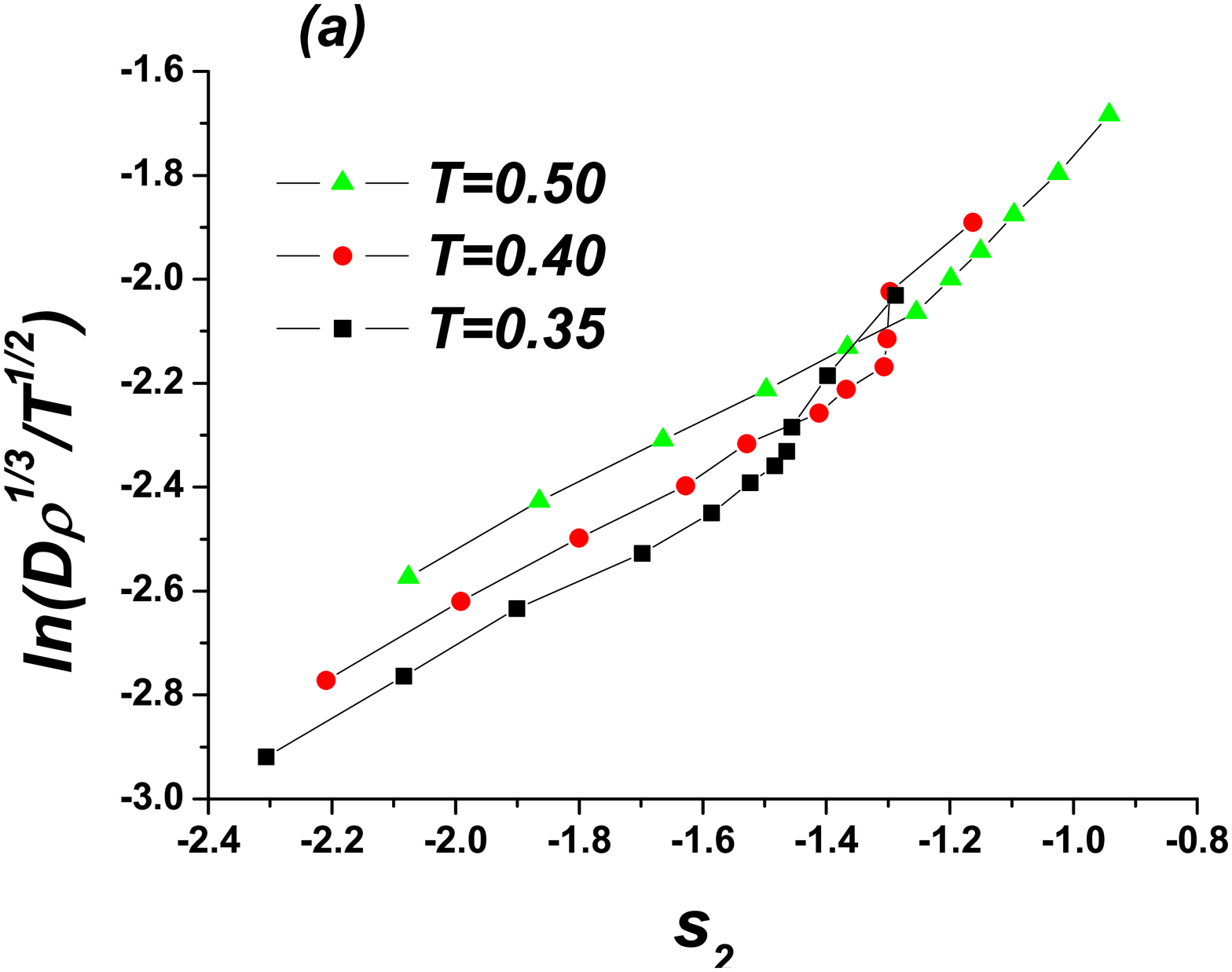}%

\includegraphics[width=8cm, height=8cm]{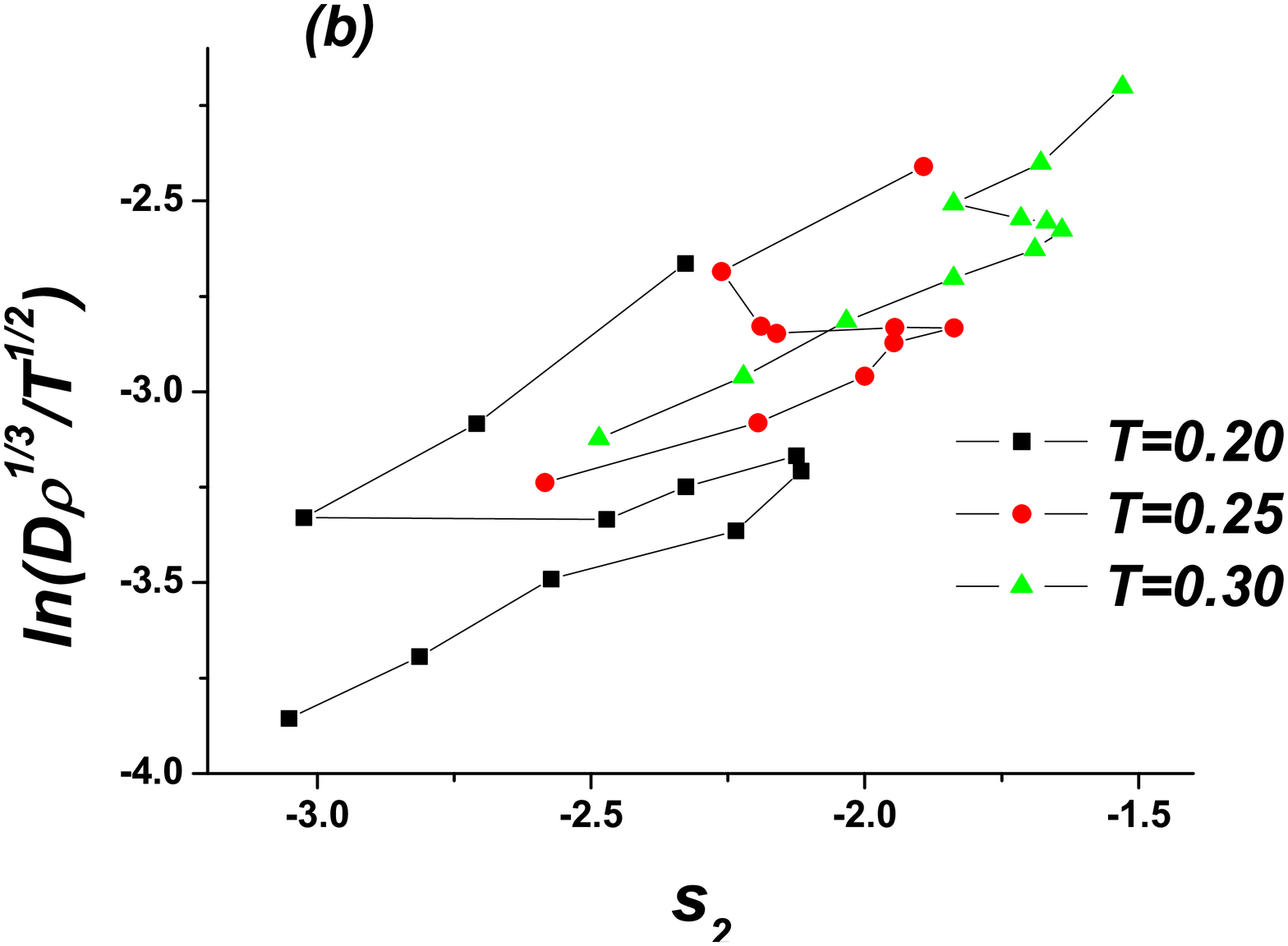}%
\caption{\label{fig:fig13} The diffusion scaling with the pair
contribution to the excess entropy at (a) high and (b) low
temperatures for the soft repulsive shoulder system.}
\end{figure}

\begin{figure}
\includegraphics[width=8cm, height=8cm]{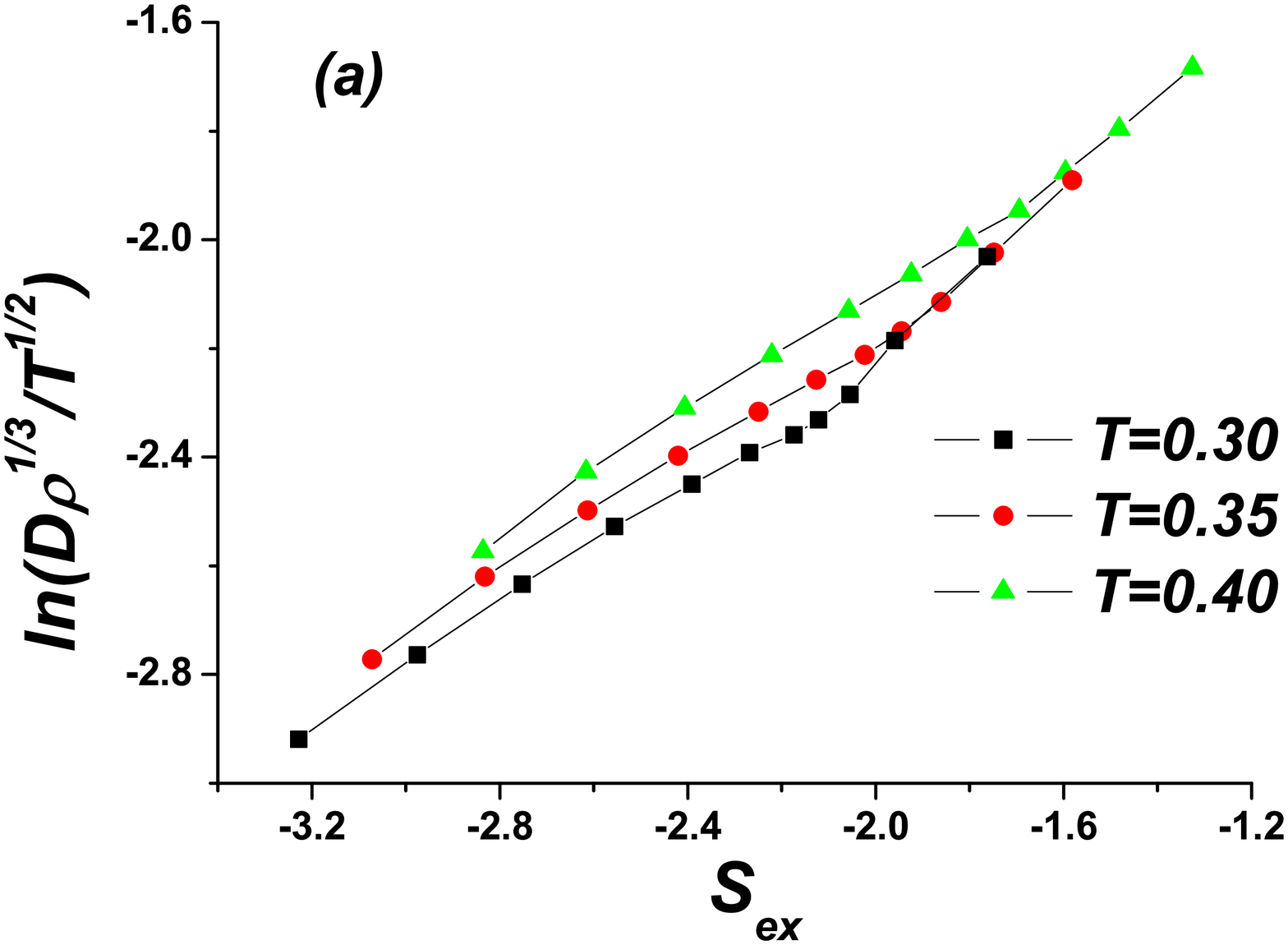}%

\includegraphics[width=8cm, height=8cm]{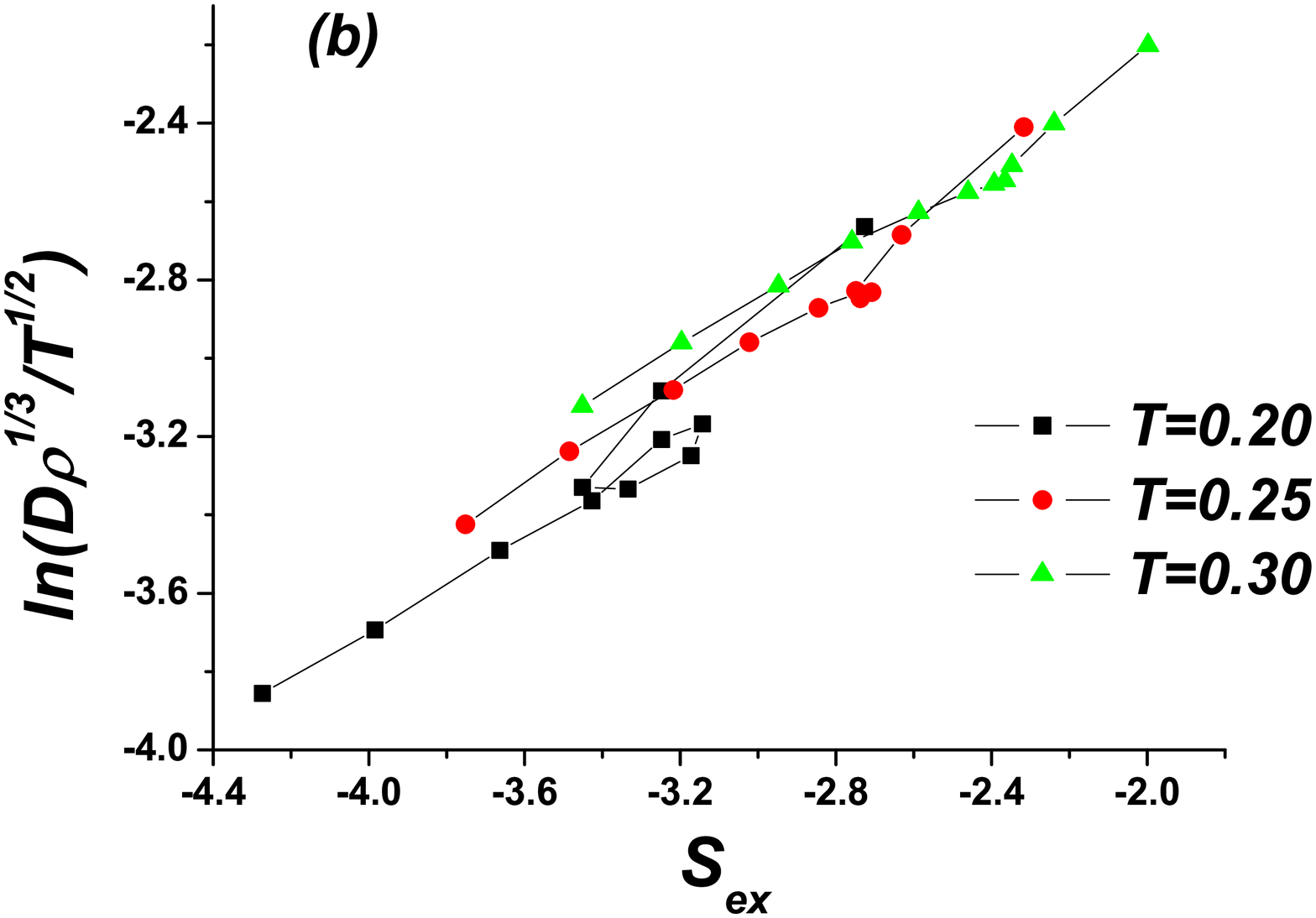}%
\caption{\label{fig:fig14} The diffusion scaling with the excess
entropy at (a) high and (b) low temperatures for the soft
repulsive shoulder system.}
\end{figure}

The diffusion scaling with the full excess entropy is shown in the
figure 14 (a) and (b). One can see from these pictures that the
scaling rule works good for the temperatures $T=0.5$ and $T=0.4$
but already for $T=0.35$ the deviation from the linear behavior
occurs. This deviation develops more as the temperature decreases.
At $T=0.25$ a self crossing loop occurs. This loop even enlarges
at lower temperatures. It is worth to note that the curve at low
temperature $T=0.2$ consists of two linear parts connected by the
self crossing loop. It means that the Rosenfeld formula is valid
in some narrow regions, but not in the whole range of densities.

In order to see the reason for the occurrence of the self crossing
loop we compare the qualitative behavior of the diffusion
coefficient and excess entropy. For this reasons we measure the
densities corresponding to the minimum and maximum of the
diffusion ($\rho_{min}^D$ and $\rho_{max}^D$ correspondingly) and
excess entropy ($\rho_{min}^S$ and $\rho_{max}^S$) at the lowest
temperature we study $T=0.2$. The values we obtain are:
$\rho_{min}^D=0.44$, $\rho_{max}^D=0.55$, $\rho_{min}^S=0.40$ and
$\rho_{max}^D=0.535$. One can see that there is a mismatch in the
location of the extremal points of these two quantities and
therefore there are some regions where the qualitative behavior of
diffusion and excess entropy is opposite. Taking into account
exponential dependence supposed in the Rosenfeld formula one can
expect that even small discrepancies in the qualitative behavior
can lead to large errors in the scaling relation.

\section{Conclusions}

This articles presents a simulation study of the applicability of
the Rosenfeld entropy scaling to the systems with negative
curvature and bounded potentials. It was shown that the excess
entropy scaling can not be applied to such systems at low enough
temperatures. Interestingly all of the systems considered here
demonstrate anomalous diffusion behavior in some regions of
temperatures and densities. It makes questionable if the Rosenfeld
relation is applicable for the systems with diffusion anomaly.
These results are in contradiction with the results of Refs.
\onlinecite{errington,mittal}. This contradiction may be
attributed to the differences of considered potentials and
simulation methods, however, this question requires further
investigation and will be a topic of a subsequent publication.

One can suppose that the excess entropy scaling is invalid for the
systems with negative curvature potentials such as repulsive
shoulder potential. A possible reason for this can be related to
the fact that such systems are effectively quasibinary
\cite{FFGRS2008,GFFR2009}. As it was mentioned in the
introduction, the original Rosenfeld idea was based on the
connection of a liquid under investigation to the effective hard
spheres liquid. At the same time liquids with negative curvature
potentials may be approximated by a mixture of hard spheres of two
different sizes. The concentration of components of such mixture
is pressure and temperature dependent. As it was shown in
literature (see, for example, \cite{dzugmix}) the excess entropy
scaling holds for binary mixtures too. But in the case of
quasibinary mixture since the effective concentration depends on
the pressure and temperature the behavior becomes more complex.
This brings to the breakdown of the scaling rules for this case.

Obviously, the systems with bounded potentials can not be
approximated by hard sphere potentials too. It seems that this may
be the reason of violation of Rosenfeld entropy scaling for these
systems.

\bigskip

\begin{acknowledgments}
We thank V. V. Brazhkin and Daan Frenkel for stimulating
discussions. Our special thanks to Prof. Ch. Chakravarty who
attracted our attention to the problems considered here. The work
was supported in part by the Russian Foundation for Basic Research
(Grant No 08-02-00781).
\end{acknowledgments}


\end{document}